\documentstyle[aps,preprint]{revtex}
\begin{document}
\draft
\title{ Parametric generation of second sound in superfluid
             helium: linear stability and nonlinear dynamics. }

\author{ Dmitry Rinberg$^1$ and Victor Steinberg}

\address{Department of Physics of Complex Systems\\
          The Weizmann Institute of Science, 76100 Rehovot, Israel\\
          $^1$ also: Bell Laboratories, Lucent Technologies, Murray Hill, NJ 07974, USA.\\}

\date{\today}
\maketitle

\begin{abstract}
We report the experimental studies of a parametric excitation of a second
sound (SS) by a first sound (FS) in a superfluid helium in a resonance cavity.
The results on several topics in this system are presented:
(i) The linear properties of the instability, namely, the threshold,
its temperature and geometrical dependencies, and the spectra of SS just
above the onset were measured.
They were found to be in a good quantitative agreement with the theory.
(ii) It was shown that the mechanism of SS amplitude saturation is
due to the nonlinear attenuation of SS via three wave interactions between
the SS waves.
Strong low frequency amplitude fluctuations of SS above the threshold
were observed.
The spectra of these fluctuations had a universal shape with
exponentially decaying tails.
Furthermore, the spectral width grew continuously with the FS amplitude.
The role of three and four wave interactions are discussed with
respect to the nonlinear SS behavior.
The first evidence of Gaussian statistics of the wave amplitudes for
the parametrically generated wave ensemble was obtained.
(iii) The experiments on simultaneous pumping of the FS and independent SS
waves revealed new effects.
Below the instability threshold, the SS phase conjugation as a result
of three-wave interactions between the FS and SS waves was observed.
Above the threshold two new effects were found: a giant amplification
of the SS wave intensity and strong resonance oscillations of the SS wave
amplitude as a function of the  FS amplitude.
Qualitative explanations of these effects are suggested.

\end{abstract}


\narrowtext

\section{\bf Introduction.}
\label{sect_Intr}

Superfluid He$^4$ exhibits rather unique nonlinear acoustic properties
particularly in the
vicinity of the superfluid transition.
As known for a long time \cite{Khal_b-65,Wilks_b-67}, it is rather  easy to
observe a
second sound  shock wave formation even for rather small amplitudes in the
close vicinity of
the superfluid transition temperature, $T_{\lambda}$
\cite{Goldner-93,David-95}.
Nonlinear interaction of first sound (FS) and second sound (SS) is another
manifestation of
nonlinear wave phenomena in the superfluid helium.

There are a few advantages that make the superfluid helium a very appropriate
system for
quantitative study of nonlinear wave dynamics.
First, parameters of nonlinearity can be easily tuned in a wide range by
changing
temperature in the vicinity of $T_{\lambda}$.
Second, thermodynamic and kinetic properties of the superfluid helium are
well known.
And third, very high precision and accuracy as well as an excellent
experimental control,
can be achieved in the superfluid helium experiment.

In a contrast to self-interaction of first and second sounds that have been
studied a lot
\cite{Goldner-93,David-95,Iznan-83}, there were only a few works dedicated
to interactions between sounds.
The process of SS generation by FS in an infinite geometry was considered
theoretically long time
ago \cite{Pushkina-74,Pokrov-76}.
Two basic mechanisms of FS to SS conversion due to three wave (3W)
interaction
were predicted: parametric generation of SS and Cherenkov radiation.
Here and further, we follow the terminology defined in the pioneer
work of Pokrovskii and Khalatnikov
\cite{Pokrov-76}.
In both cases the finite
amplitude SS waves are generated above some
threshold amplitude of a FS wave as a result of an instability.
In spite of several attempts made during the years\cite{priv}, an experimental
verification of both instabilities
was absent.
The only experiment dealing with a nonlinear interaction between two sounds
was
conducted by S. Garrett et al.
\cite{Garrett-78} more than 20 years ago.
These authors observed the process of the nonlinear conversion of SS into
FS, that was an inverse process to
one, considered in Ref.\cite{Pokrov-76}.
This observation gave an indication of the existence of the nonlinear
interaction between two acoustic
branches in the superfluid helium but left aside the problem of the
instabilities due to two mechanisms predicted
theoretically.

A nonlinear interaction of different wave branches is a rather common
phenomenon in condensed
matter physics.
Some of the well known examples are spin waves in ferro- and
antiferromagnets interacting
with microwaves \cite{L'vov_b-94}, Langmuir waves  in a plasma interacting
with electromagnetic waves, two sound branches in a dusty  plasma.
There are numerous examples of interactions between sound waves and  various
types of
collective oscillations in solids.
Some of them are a phonon-plasmon interaction in piezoelectric
semiconductors, interaction
either longitudinal or shear sounds with  electromagnetic waves in
piezoelectrics or
magnets, an interaction of sound and  surface capillary waves.
In all these systems, externally driven waves at a sufficiently large
amplitude  become
unstable with respect to generation of waves of another type
\cite{L'vov_b-94,Cross-93}.
Only in a very few systems mentioned above, these instabilities were
experimentally observed.

This paper is dedicated to one particular example of a wave interaction
- a parametric instability.
In parametrically driven systems, a driving field excites  pairs of
waves propagating in almost opposite
direction and having approximately a half of a pumping frequency.
The threshold of the parametric instability is  defined by the
balance between an energy, transferred from a driving field to a
wave  ensemble, and a wave attenuation.
At the instability threshold the wave attenuation is compensated by
the pumping.
Above the instability onset, the wave amplitude should grow
exponentially, but saturates due to
nonlinear interactions different from one that causes the instability.

The wave dynamics above the onset is defined by nonlinear wave interactions
between excited
waves.
In the most cases amplitudes of parametrically generated waves are
sufficiently low, so that
only the lowest orders in wave interaction processes,  three-wave (3W) and
four-wave (4W)
interactions, should be taken into account.
The wave systems can be divided into two major classes according to  the
type of their
dispersion law.
The first one is the decay type, which allows 3W interaction (for a power
type dispersion
law, this gives a criteria: $\partial^2\omega/\partial k^2>0$ ).
In such systems, 3W processes are responsible for the amplitude saturation
of the parametrically
excited waves \cite{Cross-93}.
These processes provide an energy transfer from  the resonance frequency
to about twice of it, at which the waves attenuate effectively.
For the second type of systems with a non-decay type of the dispersion relation
($\partial^2\omega/\partial k^2<0$), 3W processes are forbidden, and 4W
processes are
responsible for the  amplitude saturation.
The saturation occurs due to "dephasing" of pairs of excited waves in respect to a
phase of a
pumping frequency \cite{L'vov_b-94}.

The parametric wave generation is observed in a wide class of nonlinear
media.
The most  prominent examples are spin waves in ferro- and
antiferromagnets driven by a microwave field,
ferrofluid surface waves subjected to an ac tangential magnetic
field, surface waves in liquid
dielectrics parametrically excited by an ac  electric field, and a
parametric excitation of surface
waves by a vertical vibration.
In the recent years, the latter system, called the Faraday crispation or ripples,
was extensively studied
due to easy visualization of patterns which appear above the
instability threshold \cite{Cross-93}.

In this paper, we present the first experimental observation of the
parametric instability of FS in the
superfluid helium.
A FS wave with a large enough amplitude is unstable in respect to
creation pairs of SS
waves propagating almost  in opposite directions and having
approximately a half of a FS frequency.
We study the statistical and dynamic nonlinear properties  of the
ensemble of the SS waves above the
instability threshold, and the interactions of the parametrically
generated SS waves with a SS wave pumped
externally.
The short reports on each of these subjects were published recently
elsewhere \cite{Rinb-96,Rinb-97,Rinb-98} and also presented in Ref.
\cite{Rinb_th-97}.

This paper is written as follows.
First, in the Section \ref{sect_Theory} we present a theoretical
background for the wave interactions in the
superfluid helium and, particularly, for the parametric instability.
The physics related to the threshold phenomena and to behavior of a SS
wave ensemble above the threshold is
described.
At the end of this section, we compare a novel experimental system
to study the parametric instability
with two well known systems: the Faraday ripples and the spin waves
(Sect.\ref{sect_T_Comp}).
The Section \ref{sect_ExpSetup} is dedicated to an experimental setup.
The next Section consists of three major parts.
In the Section \ref{sect_R_Lin}, the results related to
the threshold phenomena: temperature dependence of the threshold, two
types of a SS spectrum above the
threshold and their temperature dependence are presented and discussed.
Next, in the Section \ref{sect_R_Nonlin}, we discuss the results related to
behavior of the SS wave ensemble above the instability onset. The
mechanism of the SS amplitude saturation, the
structure of a wave packet above the threshold, its dependence on FS
amplitude and temperature, and
statistics of the interacting SS waves are considered.
The separate Section, (Sect.\ref{sect_R_SSpump}), is dedicated to
experiments on interaction between an
independently generated SS wave with a FS-SS wave ensemble below and
above the parametric instability
threshold.
Here, we present our results on a SS phase conjugation below the onset
of instability, a giant
SS amplification and strong amplitude fluctuations above the onset.
The last section (Sect.\ref{sect_Concl}) concludes the paper.

\section {\bf Theoretical background.}
\label{sect_Theory}

\subsection{\bf Wave interaction in superfluid helium}

As was shown in Ref. \cite {Pokrov-76}, an effective way to study
nonlinear waves is to use a Hamiltonian
formalism.  The equations of two-fluid  hydrodynamics in the
Hamiltonian form were first written down in
Ref.  \cite{Pokrov-76b}.

Linear properties of waves defined by quadratic terms in Hamiltonian:
\begin{equation}
     H_2 = \sum_{\bf{K}} \Omega_{\bf{K}} a_{\bf{K}} a_{\bf{K}}^{\star} +
               \sum_{\bf{k}} \omega_{\bf{k}} b_{\bf{k}} b_{\bf{k}}^{\star},
\label{eq_ham2}
\end{equation}
and, particularly, by the dispersion relations for FS and SS: $\Omega = c_1
K$ and $\omega = c_2 k$, respectively.
Here and further, $a$ and $b$ are the wave amplitudes of FS and SS,
$c_1$ and $c_2$ are the FS and SS sound velocities,
capital letters ($\Omega$, ${\bf K}$) correspond to the FS  frequency and
$k$-vectors,  and small letters
($\omega$, ${\bf k}$) to SS.

The next term in Hamiltonian, $H_3$, is responsible for three wave
(3W) interactions.
This term is essential only in systems where a dispersion relation
allows 3W interaction processes.
The most general $H_3$ term can be written as
\begin{equation}
   H_3 = \sum_{\bf{k_1,k_2,k_3}} \left( V_{\bf{k_1,k_2,k_3}}
a_{\bf{k_1}} a_{\bf{k_2}}^{\star}
              a_{\bf{k_2}}^{\star} \delta(\bf{k_1} - \bf{k_2} -
\bf{k_3}) + c.c. \right),
\label{eq_ham3}
\end{equation}
where $V_{\bf{k_1,k_2,k_3}}$ is the matrix element of the 3W interaction.
For particular case of waves in superfluid helium, four different
types of 3W interaction should be
considered.
Two of them are self-interactions of FS and SS.
The following terms in the Hamiltonian describes these processes: $a
a^{\star} a^{\star}$ and $b b^{\star}
b^{\star}$ and their complex conjugated terms, corresponding to the
inverse processes.
Two others are interactions between different branches, FS
and SS: $a b^{\star} b^{\star}$ and
$a a^{\star} b^{\star}$ and their complex conjugated.
These interactions were considered in Ref. \cite{Pokrov-76} and coined
as the parametric decay and the Cherenkov
emission.

Both  sounds have a linear dispersion law that makes
interaction very peculiar.
The linear dispersion law is a marginal case between the decay type
($\partial^2\omega / \partial k^2 > 0$) and the
nondecay type ($\partial^2\omega / \partial k^2 < 0$)of spectrum.
For the decay type, 3W processes are permitted, while for the
non-decay type they are forbidden.
The linear dispersion law does allow the 3W resonant interaction but only for
waves propagating along one line
that strongly reduces a phase volume of possible interacting waves.

Since the self-interaction of FS is rather weak, the shock waves were
observed at rather  high incident
amplitudes \cite{Iznan-83}. Contrary to this, in the temperature range near superfluid transition,
SS velocity approaches zero, the coefficient of SS self-interaction diverges, and
the SS shock wave can be easily generated even at a rather low wave amplitude.
This strongly attracted researchers in nonlinear wave dynamics
\cite{Goldner-93,David-95,Nemir-90}.

However, the problem of the nonlinear interactions between many excited SS waves
have not been considered.
One of the experimental realizations of a multiwave paradigm is an
excitation of an ensemble  of SS waves by FS via either its parametric
decay or Cherenkov emission.

Both of later processes are the decay of one primary wave into two
secondary waves.
One FS wave in the Cherenkov process generates one FS and one SS waves,
and in the parametric decay -- two SS
waves.
In order for such a process to occur, an initial wave should
interact with a background noise and
amplifies  those particular secondary waves from the noise, which
frequency and $k$-vectors satisfy
resonance  conditions. The energy transfer from a primary wave is
proportional to a primary wave amplitude
and a matrix element of interaction, $Va$, and it should exceed the
energy attenuated by secondary waves.
The threshold amplitude of the decay process is defined by the energy balance
between pumping and attenuation.
Above the threshold, the mechanisms that are responsible for a secondary
wave amplitude saturation, are different
from those that generated these waves.
These mechanisms need special consideration and will be discussed
later for the parametric decay.

Both of these processes have inverse processes of two wave conversion into
one wave.
Inverse processes do not deal with a background noise wave amplification,
so they have no
threshold behavior.
One of the inverse processes, generation of a double frequency FS wave with a double
frequency by two SS waves, was experimentally studied by S. Garrett et al.
\cite{Garrett-78}.

\subsubsection {\bf Cherenkov radiation of SS.}
\label{sect_T_Cher}

The Cherenkov emission results from the resonance interaction between a FS
wave with the frequency $\Omega$
and the wavevector $\bf{K}$ with a pair of waves: a SS wave ($\omega'$,
$\bf{k}'$) and a FS wave
($\Omega'$, $\bf{K}'$), satisfying the following resonance conditions
(see vector diagram in Fig.\ref{fig_Vect}.a ):
\begin{equation}
     \Omega=\omega' +\Omega',
     \hspace {.5in} {\bf K=k' +K'}.
\label{eq_Cher}
\end{equation}
Here $\Omega = c_1 K$, $\omega = c_2 k$ are related to the FS and SS
waves, respectively.
Due to large difference in FS and SS velocities, particularly near
the superfluid transition,
$\eta = c_2/c_1 \ll 1$, the generated SS wave propagates almost in
the forward direction,
having the frequency $\omega' = 2\Omega\eta\sin(\chi/2)$,
and the secondary FS wave propagates in the backward direction with
almost the same
frequency and the wavenumber, as the incident FS wave.

The strength of the Cherenkov interaction which is defined by the
matrix element $V(\Omega,\chi)$,
was calculated in Ref. \cite{Pokrov-76}.
The FS threshold amplitude for the Cherenkov radiation is defined as
\begin{equation}
      a_{th}^{\infty} =
\frac{\sqrt{\gamma_1(\Omega')\gamma_2(\omega')}}{V(\Omega,\chi)}.
\label{eq_CherThresh}
\end{equation}
It has minimum at $\chi=\pi$, and is scaled with the frequency as
$a_{th}^{\infty}\sim \sqrt{\Omega}$.
Here $\gamma_1$ and $\gamma_2$ are the attenuation rates of the FS and SS
waves, respectively.

\subsubsection {\bf Parametric decay instability.}
\label{sect_T_ParInst}

Another 3W resonance process between the FS and SS waves is the  parametric
decay of one FS
wave into two SS waves:
\begin{equation}
     \Omega=\omega_1 +\omega_2, \hspace{.5in} {\bf K=k_1+k_2}.
\label{eq_ParamDec}
\end{equation}
Due to smallness of the ratio $\eta$, the parametrically generated SS
waves have almost a half
the FS frequency, $\omega_{1,2}\approx\Omega/2$, and propagate in
almost opposite directions,
${\bf k_1\approx -k_2}$.
The vector diagram of this process is shown in Fig.\ref{fig_Vect}.b.
A more detailed analysis of the resonance conditions reveals an
angular dependence of the SS frequency
\begin{equation}
     \omega_{1,2}=\frac{\Omega}{2}\left( 1\pm \eta\cos\theta \right),
\label{eq_omega12}
\end{equation}
where $\theta$ is the angle between the plane of the SS wave propagation direction
and the FS wave vector, see Fig.\ref{fig_Vect}.b.
In $k$-space, the resonance surface of SS waves interacting with the FS
wave, is the ellipsoid of revolution
(almost a sphere).
The matrix  element of the parametric decay process has been also
derived in Ref. \cite{Pokrov-76} and has the following
form
\begin{equation}
U(\Omega,\theta)=\sqrt{\frac{\Omega^3}{32\rho}}c_1^{-1}(U_1+U_2-2\cos^2\theta),
\label{eq_ParMatrEl}
\end{equation}
where $\rho$ is the density of a liquid helium, and $U_1$ and $U_2$ are
the thermodynamic functions of
temperature, pressure, density, and superfluid density:
\begin{eqnarray}
U_1 &=&  -\frac{\rho}{J\phi}(\frac{C_P}{T}\frac{\partial\psi}{\partial P}+
            \frac{\alpha_P}{\rho}\frac{\partial\phi}{\partial T}),  \\
U_2 &=&  \rho^{-1}( \frac{\partial\psi}{\partial T}+
            \frac{\alpha_P}{\kappa_T}\frac{\partial \psi}{\partial P}).
\end{eqnarray}
Here the following notations are introduced:
$\psi = \rho \alpha_P/J$,
$J=\rho\kappa_T C_P/T-\alpha_P^2$,
$\sigma=S/\rho$, $\phi = \rho_s/\rho$, where
$\rho$ is the helium density,
$\rho_s$ is the superfluid density,
$\alpha_P=-\rho^{-1}(\partial \rho/\partial T)_P$ is the isobaric thermal
expansion,
$\kappa_T = \rho^{-1}(\partial \rho/\partial p)_T$ is the isothermal
compressibility,
$C_P=T(\partial \sigma/\partial T)_P$ is the specific heat,
$S$ is the entropy per unit mass.

A FS wave of an amplitude $a$ generates SS waves at the rate $|aU|$.
The latter dissipate at the rate $\gamma_2$.
The parametric instability occurs when the SS amplification exceeds
the dissipation.
Then the FS threshold amplitude in an infinite geometry is defined as
\begin{equation}
a_{th}^{\infty}=\frac{\gamma_2(\Omega/2)}{U(\Omega,\theta)},
\end{equation}
The function $a_{th}^{\infty}(\theta)$ has a very shallow minimum for
a symmetric decay at $\theta=\pi/2$, when two
SS  waves have exactly a half the FS pumping frequency.

\subsubsection {\bf Experimental conditions for observation of both
instabilities.}
\label{sect_T_ExpCond}

According to the estimates made  in Ref.\cite{Pokrov-76}, the threshold
intensities
of the FS for both processes differ substantially.
Frequencies of the SS waves excited in these processes also differ greatly.
In the Cherenkov emission, the frequency of the excited SS waves is of
the order of $2\Omega (c_2/c_1)$,
and in the parametric decay the SS frequency is equal to $\sim\Omega/2$.
As a result, it was found in Ref.\cite{Pokrov-76}, that in an
infinite system the temperature range, where
the parametric excitation precedes the Cherenkov emission, is located
between 0.9 and 1.2 K.
However, as it will be shown later, the corresponding dissipation
length for the SS waves is too large in this
region, and any experimental cell cannot be considered as an infinite one.
The finite size effect drastically alters the threshold value of
the FS amplitude and makes impossible
the observation of the instabilities.
So, from this point of view, the vicinity of the superfluid
transition is only possible temperature range
where the effect can be observed.

Our estimates that took into account both regular and singular parts
of the thermodynamic and kinetic properties of the superfluid helium near
$T_{\lambda}$\cite{Ahlers-76} showed that the threshold value
of the parametric instability was almost independent of $\tau$ in the range of the
reduced temperatures from $10^{-6}$ to $10^{-2}$.
For the Cherenkov instability the threshold had dependence $\tau^{-0.67}$ in the same
temperature range and was much higher then that of the parametric instability.
Thus, the parametric instability  undoubtfully precedes the
Cherenkov radiation in the vicinity of $T_{\lambda}$.

The advantage to work in the vicinity of $\lambda$-point manifests in a
possibility to deal with the SS
waves with the attenuation length, $l$, that is less or comparable to the
cell size.
The attenuation length of the SS varies drastically in this
temperature range \cite{Mehrotra-84}.
This makes consideration of finite size effects very important.

\subsection{Parametric Instability: threshold related phenomena}

 From here, the rest of the paper will be dedicated to the parametric
excitation of the SS waves.
In order to create a large enough amplitude of FS excitation, the
resonant cavity cell was built  (see
Sect.\ref{sect_E_Cells}).
SS propagates mostly in the direction perpendicular to a FS $k$-vector,
along the planes of FS transducers.
So, two effects due to a cell size and boundary conditions should be
considered.
One is a finite cell size effect in the direction of SS propagation,
another is a resonant effect due to reflections
from the cavity walls in the perpendicular direction.

\subsubsection{\bf Finite size effect.}

Since  a horizontal size of experimental cells used in our experiments, is
comparable with the
attenuation length of the SS waves, the lateral boundary conditions should
be  taken into
account to define the FS threshold amplitude for the parametric instability.
A straightforward generalization of the envelope method
\cite{L'vov-77,Cherep-95} for a system of a
size $L$, where waves arrive and reflect normally to the boundaries,
with an arbitrary reflection coefficient
$r$, yields the following expression for the threshold \cite{Rinb-96}:
\begin{equation}
     a_{th} = a_{th}^{\infty}[1+\xi^2(l/L)^2]^{1/2},
\label{eq_ThreshFin}
\end{equation}
where $\xi$ is the minimal positive root of the equation
\begin{equation}
    \tan\xi=-\frac{(1-r^2)\xi(l/L)}{1+r^2-2r[1+\xi^2(l/L)^2]^{1/2}}.
\end{equation}
Here $a_{th}^{\infty}$ is the threshold value  for an infinite cell.
For $L < l$, the value of $a_{th}$ (along its temperature dependence)
is mainly determined by the second
term in the parenthesis in Eq.(\ref{eq_ThreshFin}), i.e., $a_{th} \sim
c_2/(LU) \propto \tau^{1.33}$ (see
curves {\bf B} in Fig.\ref{fig_Thresh} ).
Note that if the reflection coefficient tends to unity, the threshold
coincides with that for an infinite system:
$r = 1$, $\xi=0$, $a_{th}=a_{th}^{\infty}$.
For zero reflection coefficient and large ratio $l/L$, $\xi$ tends to
$\pi/2$, and one obtains
$a_{th}\approx a_{th}^{\infty}\cdot \pi/2 \cdot l/L$.

\subsubsection{\bf Resonance cavity effects.}
\label{sect_T_ResCav}

As we already mentioned, in order to increase the attainable value of
the FS amplitude in the cell, the
resonance cavity for the FS has been used.
Thus, the SS waves are generated by the FS standing wave.
One can consider two parametric excitation processes caused by two
components of the FS standing wave,
with the wave vectors $\pm {\bf K}$ (Fig.\ref{fig_ResVect}).
The momentum conservation conditions for these processes are
\begin{equation}
           {\bf K=k_1+k_2, \hspace {0.5in} -K=k_3+k_4}.
\end{equation}
The processes do not interfere unless the same SS phonon participates
in the both of them.
If it occurs, e.g., at ${\bf k_1\equiv k_3}$, the both components of
the FS standing wave contribute to the
excitation process.
Therefore this process has a lower threshold \cite{Murat-97}:
$a_{th}=a_{th}^{\infty}/\sqrt{2}$.
The frequencies of the generated SS waves in this case are
\begin{equation}
          \omega=\frac{\Omega}{2}(1\pm \eta^2).
\end{equation}

These both frequencies can be observed only if the frequency
splitting between two SS waves,
$\omega_1 - \omega_2 = \Omega \eta^2$, is larger than the SS attenuation,
i.e.,
$\Omega\eta^2>\gamma_2$.
The SS waves with wave vectors $\bf k_2$ and $\bf k_4$ interact with
different components of the
FS standing wave, $+\bf K$ and $-\bf K$.
In the opposite case, $\Omega\eta^2 < \gamma_2$, i.e., the frequency
splitting is less than the SS
attenuation, each SS wave interacts with both FS waves.
The frequencies of all SS waves must be equal : $\omega_{1,2}=\Omega/2$.
This also reduces the threshold amplitude of FS by a factor of two,
compared with that in an infinite
cell \cite{Murat-97}:
\begin{equation}
          a_{th}=a_{th}^{\infty}/2.
\end{equation}

In general, the threshold of the parametric excitation of the SS waves by
the FS standing wave in a finite
horizontal size resonance cavity for FS waves, can be written in the
following way
\cite{Rinb-96,Rinb_th-97,Murat-97}:
\begin{equation}
       a_{th} =
a_{th}^{\infty}\zeta\left(\frac{\Omega\eta^2}{\gamma_2}\right)
                \sqrt{1 + \left[\xi\left(r,\frac{l}{L}\right)\frac{l}{L}\right]^2},
\label{eq_Ath}
\end{equation}
where $\zeta$ is the numerical function of the ratio
$\Omega\eta^2/\gamma_2$ with limits:
\begin{eqnarray}
        \Omega\eta^2/\gamma_2  \ll 1,  &  \hspace {0.5in} \zeta = 1/2,
\nonumber\\
         \Omega\eta^2/\gamma_2 \gg 1, &\hspace {0.5in} \zeta
=1/\sqrt{2}. \nonumber
\end{eqnarray}

\subsubsection{\bf Splitting of SS waves spectra.}
\label{sect_T_SpSplit}

The resonant cavity effect leads to the discreteness of the SS wave
vector in the direction of a FS wave
propagation and to another interesting phenomenon, namely, the existence
of two types of spectra of the SS
waves and the transition between them as a function of closeness to
$T_{\lambda}$.
Far away from $T_{\lambda}$, two equidistant peaks around $\Omega/2$
should be observed, while closer
to $T_{\lambda}$, a single sharp peak at exactly $\Omega/2$ frequency
should appear.
Indeed, as we already mentioned in the previous subsection, the
frequency splitting between two SS waves,
generated by two components of the FS standing waves in the resonance
cavity, should be compared with
the SS attenuation.
Far away from $T_{\lambda}$, one has $\Omega\eta^2 > \gamma_2$, and
two discrete lines in the
spectra are present.
In the opposite case, $\Omega\eta^2<\gamma_2$, when the SS
attenuation is larger than the spectrum
splitting, just one line at exactly $\Omega/2$ is found.
Thus the discreteness of the resonance states is smeared out closer
to $T_{\lambda}$.
 From the equality $\Omega\eta^2 = \gamma_2$ the transition
temperature from one-peak to two peaks
spectra is defined.

\subsection{Properties of SS waves above the threshold}

\subsubsection{\bf Nonlinear amplitude saturation mechanism.}
\label{sect_T_NonlinSatur}

Recently A. Muratov \cite{Murat-97} has published a theory, based on
the Hamiltonian approach, which
describes nonlinear properties of an ensemble of weakly interacting
SS waves above the parametric
instability threshold. Let us review these results.
The crucial point in the understanding the nonlinear behavior of the
weakly interacting SS waves, is the nature of
a nonlinear saturation mechanism of the instability.
First, it was shown \cite{Murat-97}, that for the SS waves where
3W interaction is permitted, the
nonlinear attenuation is significantly more important than the
renormalization of a pumping field.
The latter mechanism was thoroughly studied in the spin wave
systems \cite{L'vov_b-94}.
In the case of the SS waves, the 3W interaction provides a very effective
channel to dissipate the energy of the
parametrically excited waves by the second harmonic generation, i.e.
generation of the SS waves with $2k$
wave vector and $2\omega$ frequency.
It becomes essential only if an angle between the wave vectors of
the interacting SS waves is smaller than
$4\sqrt{\gamma_2/\omega}$.
Indeed, the 3W interaction couples a linearly unstable mode ($k$,
$\omega$), which is amplified by the FS pumping, with a linearly
stable mode ($2k$, $2\omega$), which is dissipative.

The theory predicts the functional relation between the total density (or
intensity), $I_2$, of the parametrically excited waves and the
reduced pumping amplitude or the control parameter  $\epsilon  = a/a_{th} - 1$ .
For the mechanism of the nonlinear attenuation considered in the Muratov's
work \cite{Murat-97}, the functional
dependence is linear:
\begin{equation}
    I_2 = g \epsilon,
\label{eq_SSampl}
\end{equation}
where, $g$ is the coefficient defined by the dimensionality and the shape of
the phase volume of the
excited waves which is determined by the symmetry of the system
(boundary conditions), by the  phase
volume of the pumping waves, and by the properties of the interaction
vertex between the pumping wave and the
parametrically generated waves.
$I_2$ is maximal for 3D wave excitation, i.e., generation of waves on a
spherical $k$-surface, and minimal
for 1D wave excitation, i.e., generation of a flat wave.
Waves with a linear dispersion law interact through 3W process only
if they propagate in almost same
direction.
So, for a higher dimensionality of a phase volume of excited waves, the
density of waves propagating in
the same direction is lower.
In a contrary, in one dimensional wave excitation case, all waves
propagates along the same direction.
So, for the higher dimensionality the energy transfer to the double frequency should
be weaker, and the intensity of the
parametrically excited waves is higher.

The proportionality of the SS intensity, $I_2$, to the control parameter,
$\epsilon$, is the
distinguish property of the nonlinear attenuation mechanism of the
parametric wave saturation.
The "dephasing" mechanism leads to the different relation: $I_2 \sim
\epsilon^{1/2}$ \cite{L'vov_b-94}.

The coefficient $g$ was calculated in Ref. \cite{Murat-97} for the
cases with different dimensionality.
As we will show further, the dimensionality of the excited SS waves in two
experimental cells differs due to different lateral boundary conditions.
In the cell I with the reflecting boundaries the SS waves were
one-dimensional, or almost flat.
While in the cell II without reflection the SS waves were, probably,
two-dimensional.
The difference in the intensity of 2D and 1D waves is a factor
$\sqrt{\gamma_0/\omega_0}$, that is rather small\cite{Murat-97}.
Since the experimentally observed intensity of the excited SS waves is much
smaller than the theoretically obtained value even for 1D case, we use
the latter case for comparison.
In this case the coefficient $g$ in the expression for the total intensity
of
the parametrically excited SS waves (see Eq.(\ref{eq_SSampl})) is
\begin{equation}
     g = c_2\omega \left(\frac{\gamma_2}{2B} \right)^2,
\label{eq_SSAmplSlope}
\end{equation}
where  $B\equiv B(\theta=0)$ is the triple vertex of the SS
interaction. The latter at the arbitrary angle $\theta$ between the SS wave
vectors in $P$ and $T$ variables looks as\cite{Murat-97}:
\begin{equation}
     B(\theta) = \sqrt{\frac{\omega^3\chi}{16\rho}}\left\{\chi^{-1}{\bf
D}(\chi)
                 + \phi^{-1}{\bf
D}(\phi)[\cos(2\theta)-2\cos\theta]\right\},
\label{eq_coefint}
\end{equation}
where $\chi=\rho\kappa_T/J$,
${\bf D}=\frac{\partial}{\partial
T}+\frac{\alpha_P}{\kappa_T}\frac{\partial}{\partial P}$, and $\psi$ and $J$
are defined after Eqs. (8) and (9).
In order to estimate the value of the intensity of the excited SS waves at a
given value of the
control parameter of the instability, $\epsilon$, one needs to evaluate the
triple vertex of
the SS waves interaction.
By simple algebra the expression for $B(\theta=0)$ can be rewritten in the
following form:
\begin{eqnarray}
     B(0) &=& B_1+B_2, \\
     B_1  &=& c_1\rho U\sqrt{\frac{2}{\psi}
             \left( \frac{\omega}{\Omega}\right)^3
             \left(\frac{\partial T} {\partial P}\right)_{\rho}},
     \label{eq_coefB1int}\\
     B_2  &=& \sqrt{\frac{\omega^3}{16\rho \psi}
              \left(\frac{\partial T}{\partial P}\right)_{\rho}}
              \frac{\rho\alpha_P}{J\phi} \frac{\partial \phi}{\partial P}
              \left[\left(\frac{\partial T}{\partial P}\right)_S^{-1}
              - \left(\frac {\partial T}{\partial
P}\right)_{\rho}^{-1}\right].
   \label{eq_coefB2int}
\end{eqnarray}

Let's estimate the triple vertex $U$ of the FS and SS wave interaction at
the reduced
temperature $\tau=3.3\times 10^{-4}$ using the expression for the threshold
intensity of FS:
\begin{equation}
     I_1 = (2\rho c_1)^{-1}(\delta P)^2 =
           c_1\Omega\left(\frac{\gamma_2}{2U}\right)^2.
\label{eq_FSint}
\end{equation}
At these conditions one finds $\gamma_2=84(sec)^{-1}$, $c_1=2.177\times 10^4
(cm/sec)$,
$C_P=56.2 (J/mol K)$, $\alpha_P=-0.065 (K^{-1})$, and the experimental value
of the threshold
pressure amplitude of the FS wave is $\delta P=50 Pa$.
Then one gets $U=3.5\times 10^5(cm/g\cdot sec)^{1/2}$, and for $B_1=10^6
(cm/g\cdot sec
)^{1/2}$ and $B_2=3.8\times 10^4 (cm/g \cdot sec )^{1/2}$.
Here we estimate $\kappa_T=1.49\times 10^{-8} (cm^2/dyn)$, $J=0.136
(K^{-2})$, $\psi=-0.07
(g\cdot K/cm^3)$,
$(\frac{\partial T}{\partial P})_{\rho}=-2.29\times 10^{-7} (K\cdot
cm^2/dyn)$,
$(\frac{\partial T}{\partial P})_S=-6.9\times 10^{-9} (K cm^2/dyn)$, and
$\frac {\partial\ln(\rho_S/\rho)}{\partial P}=-8.3\times 10^{-6}
(cm^2/dyn)$.
Thus, finally one obtains the following value for the coefficient:
$$ g = 4.5\times10^{-3}(erg/cm^2 sec) = 4.5\times 10^{-4}(\mu W/cm^2)=
4.5\times 10^{-6}(W/m^2).$$
The latter one should be compared at, e.g.,  $\epsilon=1$ with $I_1=0.04
(W/m^2)$ at the same
conditions.

As we have already pointed out the nonlinear attenuation is proportional to
the
strength of the interaction between the SS waves.
The same type of interaction is responsible for the creation of the SS shock
waves which
strength increases strongly while  approaching the superfluid transition.
Thus one can expect that the nonlinear attenuation can increase as well near
$T_{\lambda}$.
Then the intensity of the SS waves parametrically generated above the
threshold, should
decrease correspondingly with the scaling  $g\sim \tau^{\alpha}$, where
$\alpha>0$.
Indeed, the estimated scaling of asymptotic temperature behavior of the
coefficient $g$
is $g\sim \tau^{0.33}$.

\subsubsection{\bf Broadening of the SS spectrum above the threshold.}
\label{sect_T_broad}

Another manifestation of the nonlinear wave interaction is a spectral
broadening of the parametrically generated waves.
In weakly dissipative nonlinear media, 3W and 4W play different roles
in wave kinetics.
Although 3W and 4W processes control the amplitude saturation of the
parametrically excited waves for the decay
and nondecay types of the dispersion law respectively,  only the 4W processes
allow the wave interactions within a
narrow frequency range around the main peak at the parametric
resonance frequency.
Therefore, the spectral shape around the peak is solely controlled by
the 4W processes.
Thus, one can expect that the spectral shape at the parametric
resonance frequency exhibits some universal
properties independent of details of the wave interaction and the
type of the dispersion law.
Wave amplitudes in different parametrically  driven systems can be
limited by different nonlinear
mechanisms,  their spectral broadening mechanism is universal.

The first experimental evidence and following up theoretical
explanation of the exponential spectra based on
a kinetic theory, were obtained for the parametrically generated spin
waves\cite{Krutsenko-78,Kotyuzh-84}.
A theoretical idea behind the explanation is very general and
convincing one \cite{Krutsenko-78}.
There are two reasons for the spectral broadening: a thermal noise
above the onset and a ''intrinsic''
noise due to 4W scattering of spin-waves.
The latter becomes important at higher values of the control parameter.

The broadening due to the thermal noise produces the squared
Lorentzian shape spectrum \cite{L'vov_b-94}.
The ''intrinsic'' noise results from the 4W interaction.
This interaction does not transfer energy far away from the region of
the parametric resonance frequency peak,
but pumps it to neighboring modes and broadens the spectrum.
In general, the wave packet shape is described by a nonlinear
integral equation for wave amplitude
correlation functions.
It was proved that such equation has universal exponentially decaying
tails in the frequency domain
\cite{Lvov-78}.
The 4W resonance interaction is present in any medium with
parametrically driven waves.
Thus, the kinetic theory predicts the universal spectral broadening
in such systems irrespectively of the wave
dispersion law.

A theory of the spectral broadening in the parametrically excited SS
waves was developed recently by A. Muratov\cite{Murat-97}.
According to Ref. \cite{Murat-97}, the thermal width of the central
line in the SS spectrum can be estimated from the following expression
\begin{equation}
    \frac{(\Delta\omega)_T}{\gamma_2} = \frac{\omega^2 \gamma_2
k_BT}{4\pi^2c_2^2I_2},
\end{equation}
where $k_B$ is the Boltzmann constant, and $I_2=g\epsilon$ is the SS
waves intensity,
Eq.(\ref{eq_SSampl}).
Using estimates for $g$ from the previous section and from
Ref.\cite{Murat-97},  at
$\epsilon=1$ one obtains $(\Delta\omega)_T/\gamma_2 \sim 10^{-6}$,
which is extremely small and unattainable in a real experiment.

There are two possible channels for the "intrinsic" noise.
The first is a direct 4W process, i.e. direct scattering of two SS waves on
two SS waves. The second is two consequent 3W
processes: Two SS waves generate a double frequency SS wave which decays into
two
new SS waves with a
frequency of the initial waves.
This scattering process can be considered also as an indirect 4W process.
As calculations show \cite{Murat-97}, such second order 3W scattering (or the
indirect 4W process)
is more effective than the direct 4W
scattering of two SS waves on two SS waves.
The "intrinsic" reduced width of the central line in the SS spectrum depends
on the dimensionality of the excited SS waves \cite{Murat-97}
\begin{equation}
    \frac{\Delta}{\gamma_2}\sim
\epsilon^{8/(9-D)}\cdot(1+\epsilon)^{\frac{D-1}{9-D}},
\label{eq_SpWidth}
\end{equation}
where $D$ is the space dimensionality of the SS waves.
Thus, at $D=2$ one has
$\Delta/\gamma_2\sim\epsilon^{8/7}\cdot(1+\epsilon)^{1/7}$, and at
$D=1$
one gets $\Delta/\gamma_2\sim \epsilon$.

\subsection {\bf Comparison of different parametrically driven wave
systems.}
\label{sect_T_Comp}

At this point, it will be instructive to compare the parametric instability
in
the  novel system with another two systems which were studied the most
extensively: the spin waves in  magnetics generated by a microwave pumping, and the
Faraday ripples excited on a fluid free surface by  vertical vibrations.
From a theoretical point of view, these two systems exhibit different  aspects
of
a nonlinear behavior of the parametrically excited waves.
The Faraday crispation is considered as a canonical example of a pattern
forming system, and was studied by using an amplitude equation dynamical
approach \cite{Cross-93}.
Whereas the spin waves exhibit very large number of interacting modes.
In this system, an irregular state of excited  waves is expected to
appear above the threshold of the parametric
instability \cite{L'vov_b-94}.
Then the statistical kinetic approach, based on a random phase
approximation,
is used to describe experimental results \cite{L'vov_b-94}.
The main distinction of the parametrically excited spin waves is an
enormous aspect ratio, $\Gamma=L/l$, which is usually 100 and more (at
$Lk\approx 10^5-10^6$) at very  minute relative dissipation rate,
$\gamma/\omega \leq 10^{-6}$.
Here $L$ is the horizontal cell size, $l$ and $\gamma$ are the
attenuation length and the dissipation rate of the waves, respectively.
The dispersion law of the spin waves in most systems is of the
non-decay type.
In this case, the 3W resonant interactions of the spin waves above the
threshold are forbidden, and only the 4W resonant interactions are possible.
As was suggested theoretically and verified experimentally, a
non-dissipative mechanism of the 4W interactions is responsible for a saturation
of
spin waves amplitude\cite{L'vov_b-94}.
This mechanism involves a phase detuning ("dephasing") between the spin  waves
and
the  electromagnetic rf pumping.
Even though, the 4W interactions is of the next order approximation in
the wave amplitude compared to the 3W case, the 4W interactions are responsible for
the amplitude saturation, when  3W processes are forbidden by the dispersion
relation \cite{Cross-93}.
Due to all these features the wave turbulent-like state is observed just above
the instability threshold.

At this point we would like to clarify  various definitions of highly irregular
wave states.
The term "wave," or "weak," turbulence was introduced to distinguish the highly irregular
state of many interacting waves from a strong hydrodynamic
turbulence\cite{L'vov_b-94,zakhar}.
A common feature of a fully developed wave and a hydrodynamic turbulence is an
observation of a wide  spectrum of excited modes in the frequency and wavenumber domains,
where the energy pumping and the dissipation have very different scales.
The major factor that makes wave (weak) turbulence different from the
hydrodynamic one is a presence of a small parameter in a theory, such as, e.g., ratio
of a wave energy to a ground state energy.
While there is no theory of hydrodynamic turbulence, the self consistent perturbation theory of
wave turbulence has been developed in the works of Zakharov's school \cite{zakhar}.
One of the strong assumptions of this theory is the randomness of the phases of waves creating
turbulent state \cite{L'vov_b-94}.\\
The word "turbulence" in a context of wave problems sometimes has different meaning.
The irregular wave state with many interacting waves can be also called turbulence.
In the case of parametric pumping, such a state has been coined as a parametric turbulence
\cite{L'vov_b-94}. In this case, in a contrast to the developed wave turbulence, the
energy pumping and the dissipation occurs on the same spatial scale.
A narrow, almost singular spectrum characterizes the parametric turbulence.
 The theory of the parametric turbulence was first developed and applied to a spin wave system \cite{L'vov_b-94}.
Thus, the parametric wave turbulence is the
irregular state of waves with random phases,
that consists one or several wave packets and does not exhibit an inertial
range with a wide range of algebraic power law
in a spectrum, like in the fully developed wave turbulence\cite{L'vov_b-94,zakhar}.\\

The parametrically generated surface waves exhibit a very different type of behavior
depending on the value of $k$-vector.
In a small $k$ limit, gravity waves with thenon-decay type spectrum
($\partial^2  \omega/\partial k^2 < 0$) are excited, while in a large $k$ limit, capillary
waves with the decay type spectrum ($\partial^2  \omega/\partial k^2 > 0$) are found
In the intermediate regime of the capillary-gravity waves, there exists the
value of $k$ at which only collinear resonant 3W interactions  are
permitted.
It is coined as the second-harmonic resonance, and the corresponding
capillary-gravity waves are called the Wilton's ripples\cite{Hammack-93}.\\
In the capillary wave regime, In the capillary wave regime, 3W resonance
conditions define the angle of strong
nonlinear interactions.
These interactions are responsible for both amplitude saturation and pattern formation
\cite{Zhang-97,Edwards-94, Chen-99}.
In a sufficiently large container, $Lk \gg 1$, the standing surface
waves above the threshold produce patterns which symmetry is found to be
independent of the container shape. However, even for the recent experiments,
where $Lk$ reaches the value 100,
for low viscousity regime ($\gamma/\omega \geq 0.01$) the aspect ratio $\Gamma = L/l$ still
remains close to unity \cite{Kudrolli-96,Binks-97}.
As the driving increases the pattern becomes time-dependent, and in
the large containers it looses its spatial
coherence via a defect nucleation \cite{Ezer-86,Tufill-89}.
Thus spatio-temporal chaotic behavior  rather than the parametric wave turbulence
was found in this system far above the threshold.

{\sl Comparison of the parametrically generated SS waves in respect to the
amplitude
saturation mechanism with  the Faraday instability.}
As discussed above the SS amplitude saturation occurs due to the nonlinear attenuation and
can be  compared with a saturation mechanism of the
parametrically driven surface waves investigated recently in details
\cite{Zhang-97,Chen-99}.
There are two major differences between these two systems.
First, the Faraday ripples have typically a considerably larger reduced
dissipation rate $\gamma/\omega$ than the parametrically generated SS waves
(at least, two-three orders of magnitude).
Second, the marginal condition for the collinear 3W interactions fulfilled
with
a great precision in the latter system for any wave frequency, due to the linear
dispersion relation, while for the Faraday instability the collinear 3W
interaction
exists only for one specific  value of the wave number in the mixed
gravity-capillary regime (the Wilton's ripples).

To derive an amplitude equation for the parametrically excited standing surface
waves, a classical multiscale method has been
applied\cite{Zhang-97,Chen-99}.
In the amplitude equation a low order  nonlinear term provides a saturation.
In a case of  spatial isotropy, the waves should be excited in any direction,
and
the mode selection occurs due to the next nonlinear terms of the form
$g_{ij}|A_j|^2A_i$, where $A_i$ and $A_j$ are the slowly varying amplitudes
of
two degenerate unstable modes.
For the known coupling  coefficients $g_{ij}$, the resulting wave pattern
can be
found from the amplitude equation \cite{Zhang-97,Chen-99}.
The derivation of the amplitude equation for the Faraday waves is greatly
simplified in the case of an ideal (inviscid) fluid, where the Hamiltonian
approach can be used, and viscous effects then can be treated
perturbatively,
analogously to the parametrically generated SS  waves\cite{Murat-97}.

For both the parametrically generated surface waves and the SS waves, a major
contribution to the wave saturation arises from the 3W interaction
\cite{Murat-97,Zhang-97,Edwards-94}. The role of triad resonance interactions
the amplitude satuartion of low viscous
surface waves was first emphasized by Edwards and Fauve \cite{Edwards-94}.
It was explicitly shown by Zhang and Vinals \cite{Zhang-97} that for these conditions
the dissipation through the excitation of the resonant stable waves at the double frequency is
the dominant channel of the amplitude  saturation.
It was also proved that 3W resonant interaction gave the major contribution into
the  coefficients of the cubic terms of the amplitude equations in this limit.

A particularly relevant case for the comparison with the SS waves is
the Wilton's ripples scenario, the parametrically generated surface waves
with the
frequency corresponding to the collinear 3W interaction.
 The nonlinear term in amplitude equation for Wilton's riples almost vanishes for all angles
between interacting waves excluding zero.
Zero angle corresponds to self interaction \cite{Chen-99}.
The characteristic minimal angle of the wave interaction depends on the dissipation rate,
$\gamma$. As $\gamma$ decreases the characteristic angle of
interactions vanishes, and the number of independent self intracting waves with the same
frequency propagating in different directions increases.
In the limit of $\gamma\rightarrow 0$, it becomes potentially infinite.

As suggested by Newell and Pomeau\cite{Newell-93} an asymptotic planform in
this case will be spatially turbulent. It means that spatial correlations of the
fields
will decay. This state was coined as the 'turbulent crystal'.
At a finite dissipation, high symmetry quasiperiodic patterns were observed
near
the Wilton's ripple condition \cite{Binks-97}.
The parametrically generated SS waves just above the instability threshold
may provide a good experimental
realization of the 'turbulent crystal' \cite{Newell-93}.

The same conclusion about the absence of a pattern formation and a
long-range
order in the  SS waves was made independently by Muratov\cite{Murat-97}.
Since only the SS waves with the wave vectors in the angle smaller than
$\Delta\theta\approx 4\sqrt{\gamma_2/\omega}\ll 1$ interact, the
long-range order at larger angles is destroyed, and so the pattern formation
is
absent\cite{Murat-97}.

This work deals with the parametrically generated SS waves in cells which
aspect ratio, $l/L$, is varied depending on the closeness to
$T_{\lambda}$, between 0.1 and 10 ($Lk$ varies  between
about 200 and 800), whereas the reduced dissipation rate is in the range
$10^{-4}<\gamma/\omega < 2\cdot 10^{-3}$.
The SS waves have the linear dispersion law which implies  $\partial^2
\omega/\partial k^2=0$.
It is a marginal case in which the 3W interactions are restricted to
the collinear waves, whereas waves propagating in different directions
interact
only via the 4W processes  which are much weaker.
All these features make this system particularly interesting to study a
highly excited state of the nonlinearly interacting SS waves.

A possibility to observe the fully developed wave turbulence in a system which has
two acoustic branches with the linear dispersion law, was discussed
theoretically
rather extensively for  the last several years, and various scalings
laws for the spectrum were predicted
\cite{Nemir-90,Pokrov-91,Kolmak-95,Kolmak-95b}.
If the nonlinear interactions between different acoustical branches are
much stronger than the wave
self-interactions inside each branch, a shock wave formation is suppressed.
Two acoustic noncollinear waves may interact efficiently via waves
from another branch.
Moreover, it was predicted that the parametric decay processes mainly
contribute into generation of an inertial range with algebraic power law
spectra.
So this system is considered as a promising candidate for an observation
and quantitative studies of the fully developed wave turbulence.
Thus, the parametric generation of SS waves may be the first step towards creation the
developed wave turbulence.
The study of this phenomena may give an insight into the nature of the multiwave
interaction in this system.

\section {\bf Experimental Setup. }
\label{sect_ExpSetup}

According to our estimates, experiments on the parametric generation
of the SS waves in  the superfluid helium should be performed in the vicinity
of
the superfluid transition with a temperature  stabilization of $\pm 1 \mu K$
or
better. Thus, we used a three-stage temperature regulated cryostat with a
helium container(experimental cell) as a third stage\cite{Stein-83,David_th-92}.
The detailed description of two experimental cells, where all
measurements were performed, follows.
There were several specifically designed technical elements that were
crucial
for success of the experiments. We will describe them below.

\subsection {\bf Cells.}
\label{sect_E_Cells}

To study the parametric instability of the FS waves in the superfluid helium
two cells of different geometries
were designed.
In both of them, we were able to (i) produce the FS wave of a sufficiently large
amplitude
with known and simple geometry, (ii)
measure the FS amplitude, (iii) to measure the SS wave amplitude, and (iv) to
test
the system by an independent SS wave.

Both cells were short cylindrical resonance cavities for FS.
Two round FS capacitive transducers form the endsides of these
cavities. Construction and calibration
techniques of the FS transducers are discussed in the
Sect.\ref{sect_E_FSTransd}. A diameter of the working part of the transducers was 50 mm.
A distance between the transducers was 3.9 mm and 2.8 mm in the
first and second cells, respectively.
Both resonance cavities had rather high quality factors of about
$Q=150$.  It allowed to obtain sufficiently large FS amplitudes at
the resonance frequencies of the cavities
and to neglect an influence of all other acoustic modes.
The first resonance frequencies for the first and the second cells
were weak functions of temperature due
to a temperature dependence of the FS velocity in the superfluid helium,
particularly in the vicinity of
$T_{\lambda}$.
In the whole working range of temperatures, the first resonance
frequencies lied in the range between
28050 and 28150 Hz and between 39380 and 39480 Hz in the first and
second cells, respectively.
The cell widths were found with a great precision by fitting our
experimental data on the temperature dependence
of the first resonance frequency in each cell using a single
adjustable parameter.
The fit was based on the known temperature dependence of the FS
velocity\cite{Maynard-76}.
The cell widths were $d=3.833$ mm and $d=2.767$ mm for the first and
second cells, respectively.

In the experimental temperature range $2\times 10^{-4}<\tau<2\times
10^{-3}$, the SS dissipation
length, $l$, varies drastically from $0.2L$ to $10L$ for frequencies
equal to a half the FS resonant frequency
for both cells.
That allowed us to study the behavior of the parametrically generated the
SS waves in different limits of
$l/L$.
The main difference in the cell design, however, was in the geometry
of bolometers and heaters for SS
wave generation and detection.

The cells were placed into a container filled with about 300 ml of
purified He$^4$ and vacuum sealed. This container had a very low
heat leak to the rest of the cryostat. A filling capillary was
disconnected by a cold valve. That was crucial for a rather high
level of temperature stabilization, below 0.1 $\mu$K. To make a
thermal contact between cell and a helium bath during the initial
cooling, the vacuum isolation space was filled with helium gas at
100 mTorr at room temperature. After the initial cooling, the
excange gas was pumped by a sorbtion pump mounted on top flange of
a vacuum can. The pump had a very weak thermal leak to the
external helium bath\cite{Stein-83}.  Such a design gave us a
possibility to cool the cell in a short time.

\subsection{\bf FS transducers}
\label{sect_E_FSTransd}

In both cells, the FS capacitive transducers similar to that described
in Ref. \cite{Heiserman_b-81}, were used.
The advantages of using the capacitive transducers are as follow: (i)
a relatively high amplitude of an
acoustic wave (in comparison to a piezoelectric), (ii)a  high
sensitivity, (iii) a flat frequency response, and (iv)
easy to build.
6 $\mu$m aluminized mylar film, stretched and glued to an external
brass ring, served as one electrode of
a capacitor.
A sandblasted brass backplate 50 mm diameter was the second electrode.
The capacitance of such a transducer was about 2 nF.
An amplitude of a driving {\it ac} voltage signal was up to 40 Vp-p,
and a {\it dc} bias up to 300 V.
The higher bias increased efficiency and sensitivity of the transducer,
but at the same time increased a
probability of an electrical breakdown.

Knowing an absolute value of the FS amplitude is very important for studying
nonlinear wave dynamics. A method of a transducer calibration using only
electrical  measurements was first developed by MacLean
in 1940 \cite{MacLean-40}; a general theory of such a calibration
can be found in Ref.
\cite{Kinsler_b-82}.
In the superfluid helium it was used in Refs.\cite{Garrett-77,Barmatz-68}.
This method is based on a few assumptions: (i) two acoustic
transducers are identical and form a resonance cavity;
(ii) all acoustic losses happen in a medium and can be measured by
the resonance characteristics of the
cavity.
While the detailed analysis is omitted, the resulting expression for the
amplitude of the acoustic pressure
oscillations at the receiver at the resonance is as follows:
\begin{equation}
  P_{2}=\left( \frac{V_{1}V_{2}}{Z}\frac{2\rho c_{1}^{2}Q}{Sd\omega
}\right)^{1/2},
\label{eq_PrCalibr}
\end{equation}
where, $V_{1}$ and $V_{2}$ are the voltage oscillations on the first
and the second transducers ( the first
transducer  is the emitter, the second is the receiver), so $V_{1}$ is
the $ac$ generator amplitude and $V_{2}$ is the
measured amplitude; $Z$ is the electrical impedance of each
transducer, $Z=1/\omega C$,  where $C$ is the
electrical capacitance; $\rho $ is the density of the medium where
acoustic waves propagate ( for helium
$\rho \approx 0.145$ g/cm$^{3}$ at $\lambda $-point); $c_{1}$ is the
sound velocity
( $c_{1}\approx 218$ m/s); $Q$ is the quality factor of the resonance
cavity, measured from the
frequency response of the cavity; $S$ is the surface area of the transducer
(
$S=19.6$ cm$^{2}$); $d$ is the
distance between transducers ( for the first cell $d\approx 3.9$ mm,
for the second: $d\approx 2.8$ mm);
and $\omega =2\pi f$ is the frequency of the sound wave.
All values, $V_{1}$, $V_{2}$ and $\omega $, correspond to the resonance.

The measurements of an acoustic sensitivity of the transducers, $M$, and
an efficiency of the acoustic
resonator, $E$, were made for both cells:
\begin{eqnarray}
   M &=&\frac{\mathrm{Measured\quad }ac\quad \mathrm{signal}\quad \left[
V_{rms}\right] }{\mathrm{\Pr essure\quad oscillations}\quad \left[
Pa_{rms}\right] }, \nonumber \\
E &=&\frac{\Pr \mathrm{essure\quad in\quad the\quad resonance}\quad \left[
Pa_{rms}\right] }{ac\quad \mathrm{voltage\quad amplitude\quad on\quad
the\quad generator}\quad \left[ V_{p-p}\right] }. \nonumber
\end{eqnarray}
It was found that $M$ and $E$ did not depend on temperature in the
experimental range and with the given
accuracy.
Temperature changes of the FS resonance frequency, of the quality factor of the
acoustic cavity (FS attenuation), of the  amplitude of the signal at the resonance
and of the helium density can be neglected.

The measured sensitivity of the transducers in the first cell was
$M_{I}=1.56\cdot 10^{-5}$ V$_{rms}$/Pa,
and in the second cell $M_{II}=1.75\cdot 10^{-5}$ V$_{rms}$/Pa.
The efficiency of the resonance cavities were: $E_{I}=3.5$
Pa/V$_{p-p}$, and $E_{II}=7.5$ Pa/V$_{p-p}$.

\subsection {\bf The First Cell.}
\label{sect_E_Cell1}

The first experimental cell (Fig.\ref{fig_Cell1}) was designed as
partially reflecting for the SS waves in a
horizontal plane.
Two SS bolometers and two heaters, evaporated on 22 mm diameter glass
substrates, were placed
diametrically opposite to each other (one bolometer is opposite to
one heater), and the two pairs
were mounted perpendicular to each other.
Each heater-bolometer pair separated by 54 mm apart, formed a
resonant, partially open cavity for the SS
waves with resonances from 12 to 40 Hz apart depending on temperature.
Possible incidental effects of these resonances on the experimental
observations will be
discussed below.
Four glass substrates covered together about half of the cell
perimeter; the remaining part was open.
Several layers of crumpled paper were put around the cell to absorb
the FS and SS waves.
To decrease the acoustic crosstalk between the FS transducers, each glass
substrate was attached to the side
of the cavity through a layer of thick filter paper.
The parametrically generated SS waves were partially reflected from
the bolometer and the heater
substrates, and partially escaped from the cell being absorbed by the
paper around the cell.

\subsubsection{\sl Flat bolometer and heater.}

Bolometers were 40 $\mu m$ wide superconducting Au-Pb stripe in the
form of a round serpentine pattern of diameter $d_b=2$ mm, evaporated
on a cover glass
(Fig.\ref{fig_BolomFlat}).
The technology of such bolometers was early developed in our
laboratory \cite{David-95,David_th-92}.
The figure of merit of the bolometers was about $\alpha =
R^{-1}(\partial R/\partial T) \approx 100$ K$^{-1}$ in a zero
magnetic field.

Since  $\lambda/d_b \ll 1$ in the working temperature range (the SS
wave length $\lambda$ changed from
0.1 till 0.5 mm), these bolometers were sensitive mainly to waves
incident almost perpendicular to the them
within the angle of $\sim \lambda/d_b$ to the normal direction, i.e.,
from 0.05 to 0.15 radian.
Thus they were able to detect the SS waves generated only in the central
part of the cell.

A heater was produced by evaporation of a thin gold film on a 22 mm
diameter glass substrate.
Contacts were made of a thick gold film( $\sim 150$ nm).
A heater resistance was 18 $\Omega$ at liquid helium temperatures.

\subsection {\bf The Second Cell.}
\label{sect_E_Cell2}

  As the experiment showed, a partial reflection of the SS waves from
the bolometers and the heaters caused
certain difficulties in an interpretation of the experimental
results and their comparison with the theory
particularly above the instability threshold.
Thus, to resolve these problems, the second cell with the non-reflecting
bolometers and the heater was designed.
The development of a new type of the bolometers and the heater was the main
technological challenge in the cell
design \cite{Rinb-00}.

\subsubsection{\sl Fiber bolometer.}

The bolometers were prepared on glass fibers which diameter was much smaller
than the SS wave
length $0.1 < \lambda < 0.5$ mm.
A superconducting gold-lead  (2:1 composition) film of an approximately
20 $nm$ thickness was
evaporated on 8 $\mu$m diameter glass fiber.
While the detailed description the bolometers can be find in Ref.
\cite{Rinb-00}, here we summarize
their main features:\\
(i) \emph{No reflections}.
The bolometers with no reflection first were proposed in Ref.\cite{Schwer-89},
where they were made by coating
glass fibers of diameter $d\approx 10\ \mu$m with a layer of gold and
tin.
However, such bolometers did not have a high enough figure of merit and
were not designed for measuring
near the superfluid transition temperature.
The diameter of the glass fibers in our case was 8 $\mu$m, close to
that reported in
\cite{Schwer-89}, and much less than a SS wave length.
A SS wave reflection from such bolometers was negligible.\\
(ii) \emph{Uniform angular sensitivity.}
Since the diameter of the fiber bolometer was much smaller than the SS
wave length, the bolometer
was sensitive to all SS waves propagating in the plane perpendicular
to its axis, in contrast to a flat
bolometer.\\
(iii)  \emph{High figure of merit. }
The figure of merit of these bolometers was $\alpha =
R^{-1}(dR/dT)\approx 30$ K$^{-1}$, that was comparable
with the best flat bolometers \cite{David_th-92}.\\
(iv)  \emph{Working temperature range is near $\lambda$-point.}
The bolometers were built on a basis of the same type of
superconducting gold-lead alloy that was used in works
 \cite{David-95,David_th-92} and \cite{Mehrotra-84}.\\
(v)  \emph{Magnetic field and current tunability.}
A capability to tune $T_{c}$ of the bolometers by varying
the bias current, while still maintaining
a high figure of merit, was essential for the use of these bolometers
in our experiments.
The experimental cell contained eight bolometers operating
simultaneously with a high sensitivity.
Since all the bolometers in the experimental cell saw the same field
from an external solenoidal magnet, it was
the individual bias current which was adjusted to bring each
bolometer's $T_{c}$ to a helium working
temperature.\\
(vi) \emph{Long time room stability.}
The superconducting layer was covered by 40 nm of MgF$_{2}$ that
decreases a rate of an oxidation at a
room temperature.
Since a technology of preparation of the bolometers and their mounting in the
cell required a high accuracy and took a
long time at a room temperature, it was necessary to prevent a fast
oxidation of the superconductor.
The bolometers could survive room temperature conditions during many
hours and almost did not change their
properties after a month being in a vacuum desiccator.

In order to have a possibility to measure an angular distribution of
the SS waves in the cell plane, eight
bolometers were mounted around the cell at equally spaced intervals
on the circumference of the cell
(Fig.\ref{fig_Cell2}).
The number of the bolometers was defined by a maximal number of
coaxial cables permitted by a given
wiring setup.
Unfortunately, only five among eight bolometers worked after cooling.

The bolometers were placed in semi-cylindrical (1.5 mm radius) groves
machined in two rings made of
G-10, mounted around the FS transducers.
Such design allowed to electrically isolate the fiber bolometers from
the FS transducer housing made of
brass, and to put them as close as possible to the working area of
the transducer membrane, so that a SS
wave path without a FS pumping was minimal.
To decrease the electrical pick-up and crosstalk by the bolometers,
all the wiring was made by coaxial cables
down to the cell and twisted pairs inside the cell.
For the same purpose, a thin copper wire (0.05 mm) was soldered
parallel to the bolometer fiber and
connected to the bolometer contact on one side of the cell.
A twisted pair was connected to other bolometer contact and this wire.
In this way an area of electrical loop which could act as an antenna
for a crosstalk, was decreased.

\subsubsection{\sl Fiber heater.}

The nonreflecting SS emitter-heater was constructed from 31 glass
fibers of 8 $\mu$m diameter
spaced 1 mm apart.
A fiber length was equal to the cell width, 3 mm.
The fiber grid was covered by $\approx80$ nm of Cr.
This grid produced a plane wave at a distance  larger than a fiber
spacing, 1 mm, and less than a grid
width, 30 mm.
The incident SS waves coming from the cell, were not reflected from the
heater.

The details of a fiber bolometer and heater fabrication, mounting,
testing and all characteristics can be
found in Ref.\cite{Rinb_th-97,Rinb-00}.

\subsection{\bf SS phase locking thermal stabilization.}

To improve a thermal stabilization of the experimental cell, a
dedicated SS phase locking (SSPL)
thermometer was built.
The idea of such thermometer was proposed and experimentally verified
by H. Davidowitz
\cite{David_th-92,David-96}.
The SS velocity is a sharp function of the closeness to $T_{\lambda}$.
So, small temperature fluctuations in the cell lead to strong
fluctuations of the phase of the SS sine
wave continuously emitted from the heater and measured by the bolometer.
The phase signal is sent to a temperature stabilization loop.

Together with the experimental cell, a separate cell for SSPL thermal
stabilization was placed inside the helium  container.
A radio frequency wave guide (cross-section $3.5\times 7$ mm$^2$,
length 60.6 mm) was used as a resonance cavity.
One side of the cavity was formed by a flat heater, evaporated on a
piece of glass substrate.
The fiber bolometer, attached to a flat G-10 plate, was mounted on
the opposite side.

The phase of the detected SS signal relatively to the phase of the
emitted signal was measured by a lock-in
amplifier.
Such a thermometer did not allow to measure the absolute value of the
temperature in the cell but
provided a very high temperature stability.
The temperature can be measured by measuring a time
propagation of the SS pulse  in the same cell.

The sensitivity of the thermometer increases while approaching
$T_{\lambda}$.
The best result on the temperature stabilization of 0.1 $\mu$K rms
was obtained at the reduced
temperature $\tau=5\times 10^{-5}$  with a SSPL frequency  $f_s=970$ Hz.

\subsection{\bf Signal acquisition.}

We developed a new technique to measure the SS spectra in a narrow
band around the central $\Omega/2$
peak.
We also describe a home-made eight-channels preamplifier with a
lock-in amplifier for a simultaneous
signal acquisition from eight bolometers.

\subsubsection{\sl Spectrum measurements.}
\label{sect_E_SpMeas}

As follows from the theory and found by our experiments, the SS
spectra should be very narrow and
centered at half of the FS frequency, $F/2 = 2\pi\Omega/2$.
For temperatures close to $T_{\lambda}$, the width of a single peak
around $F/2$,  $\Delta$,
should be much less than the SS attenuation, $\Delta/\gamma_2 \ll 1$,
where $\gamma_2 \sim 100$ Hz.
And for temperatures far from $T_{\lambda}$, the SS spectra should
consist of two peaks with a distance apart of
$\delta f = \left(c_2/c_1 \right)^2 F/2$, that is of the order of a few
Hertz.

Thus, an idea of the measurement technique is to measure a signal from
the
bolometer by a lock-in amplifier with a reference frequency exactly equal to a half  the
FS
frequency, $F_{p}/2$, to record time dependent signals from a lock-in output
and
then by, calculating their Fourier spectrum, to rebuild the real spectra in a
narrow bandwidth around $F_{p}/2$.
The measuring spectral width should be smaller than an inverse lock-in
integration time constant.
Then as shown in Ref.\cite{Rinb_th-97}, the relation between the  Fourier
component of a real, time-dependent signal at the lock-in amplifier input,
$A(\omega_0 + \delta)$, and the Fourier component
of a complex, slowly changing signal at the lock-in amplifier output,
$Z(\delta)$, is given by the following
expression at the gain equal one \cite{Rinb_th-97}:
\begin{equation}
     Z(\delta)=\frac{A(\omega_0+\delta)}{\sqrt{1+(\delta
t)^2}}\exp^{i\arctan(\delta t)},
\label{eq_lockin}
\end{equation}
where $\delta = \omega-\omega_0$ and $t$ is the lock-in
amplifier integration time.
Thus, both the phase shift and the amplitudes of spectral components of
the complex output signal should be
corrected according to Eq.(\ref{eq_lockin}).

Both components of the lock-in complex output signal were sampled and
digitized at a rate $F_{samp}$
during an interval  $T_{meas}$.
As a rule we used $F_{samp}=16$ Hz and $T_{meas}=128$ sec.
Then the complex Fourier transform of both components of the signal
provided the SS spectrum in a narrow
window, $\pm 8$ Hz, around $F/2$ with resolution of 1/128 Hz.

\subsubsection{\sl Multichannel lock-in and preamplifier.}

For the second experiment with 8 bolometers, a dedicated low noise
preamplifier and a lock-in amplifier were
built.
The low noise, wide bandwidth preamplifier with constant gains (x$10^3$)
had nine channels: eight for the cell
bolometers and the ninth for the SSPL bolometer.
A noise level in a frequency range from 100 Hz to 100 kHz did not
exceed 2 nV Hz$^{-1/2}$.
It was found that a main noise source was ratherthe  bolometers than
the preamplifier.
The preamplifier had also an adjustable simple $dc$ current source (50 -
150 $\mu$A), for a bolometer biasing.
The preamplifier was located very close to the cryostat terminal box and
had an independent battery power source
($\pm$12 V).

Signals from 8 preamplifiers were fed to a home-made 8-channel
lock-in amplifier.
The lock-in amplifier had a reference at a double frequency
compared
to the frequency of the measuring signal.
This unique property was very convenient for detection of the SS signal at a half the
FS frequency.
The input signal was multiplied by two square waves at a half the FS frequency
with
phases shifted on $\pi/2$ and integrated during a given time.
The two shifted square waves were easy to generate out of the  reference
square
wave signal taken from the \emph{SYNC} output of the FS generator, by
logical TTL
manipulation.
Low frequency signals from sixteen lock-in outputs were send to a A/D
computer card.

\subsubsection{\sl Instrumental setup.}

A general scheme of our measurements is shown in Fig.\ref{fig_InstrSet}.
The FS transducer was pumped by an $ac$ signal from a function
generator ({\bf G1}) \cite{HP3325B}.
The generator can give a sine wave with amplitude up to 40 V$_{p-p}$.
The FS amplitude was measured by an Ithaco lock-in amplifier
\cite{Ithaco} referenced from the
\emph{''SYNC''} output of the generator.
The amplitude reading was digitized and sent to the computer via a
Keithley 197 DMM ({\bf V}).
The SS measuring circuits required a biasing current of about 100 $\mu$A.
The dedicated 8-channel preamplifier with the biasing current was
designed and built.
Each bolometer was connected to the preamplifier box by a single
coaxial cable which transferred both the
bias current and the measuring signal.
All coaxial cables were grounded at a single point on the cell level.

The output signals from the preamplifiers were sent to the 8-channel
lock-in amplifier, that takes its
reference signal from the FS generator.
The lock-in amplifier was designed to have the reference frequency
twice larger than the frequency of the
signal of interest.
So to measure the parametric instability signal near a half the FS
frequency, the signal from
\emph{"SYNC"} output of the FS signal generator was directly fed to
the lock-in \emph{''Reference''}
input.
In the first experiment a standard lock-in amplifier was used.
Such a lock-in required a reference signal equal to a half the FS frequency.
A small home-made frequency divider was introduced between the generator
\emph{''SYNC''} output and the
lock-in \emph{''Reference''} channel.

For the experiments on a response of the interacting wave system to an independent SS wave,
and for a  bolometer check,
independent SS emitters-heaters were mounted in the cell.
The heater was connected to a Synthesized Function generator DS345
\cite{DS345} ({\bf G2}) that
can produce both a burst signal for measuring pulse arrival time and
a continuous sine wave.
Quartz clocks of both generators {\bf G1} and {\bf G2} were
synchronized.
This option was essential for the experiments on simultaneous pumping of the
FS and SS waves when a SS wave with a
frequency shifted from a half  the FS frequency needs to be emitted.

\section{\bf Experimental Results and Comparison with Theory.}
\label{sect_Results}

Presentation and discussion of the experimental results are divided into
three
major sections:
(i) linear properties of the parametric instability; here we discuss
all  threshold related phenomena: the temperature
dependence of the threshold and finite size effects, and the splitting of the
SS spectrum just above the threshold;
(ii) nonlinear behavior of the parametrically excites SS waves above the
threshold: the SS amplitude saturation  mechanism,
the broadening of the SS spectrum above the threshold, and the statistics of
the SSwave amplitudes;
and (iii) the experiments on a simultaneous FS and SS pumping that
include an observation of an acoustic phase conjugation
and a giant SS amplification above the threshold.

To avoid further misunderstanding we need to comment  words
"linear" and "nonlinear".
All our results are about the wave interactions in the superfluid helium that
are in general {\it nonlinear} phenomena,
however, when we use the word "linear" we mean linear in the
respect to the SS wave amplitude.
Correspondingly we use the word "nonlinear." {\it Nonlinear
phenomena}, by our classifications,  are those
where the interactions between SS waves play a crucial role.

\subsection{\bf Parametric instability: linear properties.}
\label{sect_R_Lin}

\subsubsection{\bf Experimental results}

{\sl Measurement of the threshold of the parametric instability.}

The experiments to define the parametric instability threshold of the FS
waves in the both cells were performed
in the same way.
At a fixed temperature, FS was generated at a frequency $\Omega$, which
was chosen
to be the first resonance frequency of the cavity.
For a given FS amplitude, a SS amplitude was measured by the bolometers.
Typical plots of the time averaged SS intensity in the vicinity of
$\omega=\Omega/2$  as a function of the
driving FS amplitude for the first and second cells are shown in
Fig.\ref{fig_Intens}.
Averaging time intervals for the first and second cells were 128 and 32
seconds, respectively.
A well defined threshold for the driving amplitude, at which the SS
intensity first exceeds twice a background
noise level, exists for each curve in Fig.\ref{fig_Intens}.

One can notice a striking difference between the SS intensity
dependences  for the first and second cells.
In the first cell above the instability threshold, the SS intensity
plot exhibits rather large fluctuations as a
function of the FS amplitude.
It is important to point out here, that the SS amplitude in this cell
also strongly fluctuates in time.
The characteristic time of these fluctuations was particularly long
near the onset where the time intervals
between isolated spikes could be as long as 1000 seconds (which was
the maximal sampling interval).
This time is much larger than all characteristic times in the problem.
These fluctuations limit the resolution of the threshold
determination and hinder quantitative studies of
the amplitude dynamics above the onset.
Nevertheless, they do not change the main features of the phenomena
discussed below.
The SS intensity plots for the second cell are more regular.

The temperature dependences of the FS threshold amplitude for both cells
are presented in Fig.\ref{fig_Thresh}.
The solid lines are theoretical curves which are plotted without any
fitting parameter.
As we already discussed in the Section \ref{sect_Theory}, there are two
important factors which should be taken into
account.
First, one should consider the generation of the SS waves by the standing FS
waves in a resonance cavity
rather than by the propagating ones.
Second, the cell size should be compared with the SS attenuation
length that drastically depends on
temperature.

{\sl SS spectra just above the threshold.}

Another result found at the instability threshold, was two types of the SS power
spectrum that were observed in different
temperature regions.
Examples of two typical SS power spectra at the reduced temperatures
$\tau=2.19\times 10^{-4}$ and
$\tau=6.52\times 10^{-4}$ in the first cell , are shown in the insets
of Fig.\ref{fig_Spectr1}.
The first spectrum has a single sharp peak exactly at frequency
$\Omega/2$, while the second one,
obtained further away from $T_{\lambda}$, exhibits two peaks equally
separated from $\Omega/2$.
The left peak at the lower frequency, $\Omega/2-\delta f$, is always
larger than the right one at
the frequency $\Omega/2+\delta f$.

The temperature dependence of the frequency peak shift is shown in
Fig.\ref{fig_Spectr1}. Closer to
$T_{\lambda}$, the spectrum has just one peak at the frequency
$\Omega/2$ that corresponds to the
symmetric decay of FS.
Further away from $T_{\lambda}$ ($\tau > 6\times 10^{-4}$), the
spectrum  shows two peaks equally
separated from $\Omega/2$.
A crossover from one type of the spectrum to another occurs in the
reduced temperature interval
$4\times 10^{-4}<\tau < 6\times 10^{-4}$, where both the central peak
and the pair of the separated peaks
coexist in the spectrum.

The two peak spectra are always asymmetric: the left peak
corresponding to the lower frequency is always higher that
the right one.

{\sl Features of the second cell spectra}

The geometry of the second cell differs from that of the first cell, and
this
causes some  differences in a SS response.
The most striking fact was that the SS
spectra in the second cell  were different on different bolometers and
had also different frequency shifts from an exact $\Omega/2$ value.
A typical SS spectrum just above the onset of the parametric
instability consists of a single instrumentally
sharp peak shifted from $\Omega/2$.
The peak shifts are different on different bolometers and depend on
temperature.
Their temperature dependences are shown in Fig.\ref{fig_FrShift}.
The bolometer positions in the cell are shown in the inset of
Fig.\ref{fig_FrShift}.
The bolometers 2 and 8, that are above the dotted line, connecting
the bolometers 3 and 7, have negative
frequency shifts.
The bolometers 4 and 5, located below the line, have positive frequency
shifts ( the frequency shift for the
bolometer 5 was positive but the SS signal was very noisy).
The bolometer 7 has almost no frequency shift.

\subsubsection{\bf Comparison between the experiment and theory.}
\label{sect_R_Comp}

In order to explain the experimental results on the temperature
dependence of the instability threshold
and the spectrum splitting as well, let's first  estimate the range
of variations of important parameters
in the problem, namely, the sound velocity ratio, $\eta=c_2/c_1$, the
ratio of the SS dissipation length
to the horizontal size of the cell, $l/L$, and the ratio of the
frequency splitting to the SS dissipation rate,
$\Omega\eta^2/\gamma_2$.

In the working temperature range, $10^{-4}<\tau<2\times 10^{-3}$,
$\eta$ varies between 0.006 and 0.019.
In the same temperature range,  $l/L$ changes from 11 till 0.2 in the
first cell, and from 5.5 till 0.1
in the second one.
The ratio $\Omega\eta^2/\gamma_2$ varies between 9 and 0.05 in the
first cell, and between 6 and 0.04 in the second
one.

{\sl Temperature dependence of the FS threshold amplitude.}

The experimental data on the temperature dependence of the FS
threshold amplitudes of the parametric
instability for both cells are presented in Fig.\ref{fig_Thresh}
together with various theoretical curves.
The general expression for $a_{th}$ from Eq.(\ref{eq_ThreshFin}) is used.
Curves {\bf A} in both plots show the lowest possible FS
threshold amplitude which corresponds to an
infinite cell with the resonance cavity factor $\zeta=1/2$.
Two conditions should be satisfied in order to apply this criterion:
(i) $l/L\ll 1$, and
(ii) $\Omega\eta^2/\gamma_2 \ll 1$.
It occurs in the close vicinity to $T_{\lambda}$. At $\tau=10^{-4}$
the ratios in the first and second
cells are the following:
$${\bf I.}  \hspace{0.5cm} l/L=0.21, \hspace {0.5cm}
\Omega\eta^2/\gamma_2=0.052$$
$${\bf II.} \hspace{0.5cm} l/L=0.10, \hspace{0.5cm}
\Omega\eta^2/\gamma_2=0.036$$

The maximal values of the FS threshold amplitude are realized in a
finite cell without SS reflections
from the boundaries, $r=0$, and with the spectrum splitting, i.e., at
$\zeta=1/\sqrt2$  - curves {\bf B} in
the both figures.
This opposite limit is reached at the conditions: (i) $l/L\gg 1$, and
(ii) $\Omega\eta^2/\gamma_2\gg 1$, which
are satisfied far from the superfluid transition.
For example, at $\tau=10^{-3}$ one finds
$${\bf I.} \hspace{0.5cm} l/L=4.8, \hspace{0.5cm}
\Omega\eta^2/\gamma_2=2.8$$
$${\bf II.}\hspace{0.5cm} l/L=2.3, \hspace{0.5cm}
\Omega\eta^2/\gamma_2=2.0$$

The transition from one regime to another occurs within the following
experimental temperature range
in the both cells:
$$ l/L = 1:  \hspace{0.25cm}  {\bf I.} \hspace{0.25cm} \tau\approx
3.1 \times 10^{-4},
\hspace{0.5cm} {\bf II.} \hspace{0.25cm} \tau\approx5.3\times 10^{-4},$$
$$ \Omega\eta^2/\gamma_2=1: \hspace{0.25cm} {\bf I.} \hspace{0.25cm}
\tau\approx 5.5\times
10^{-4}, \hspace{0.25cm}  {\bf II.} \hspace{0.25cm}\tau\approx
6.8\times 10^{-4}.$$
The experimental data for the FS threshold amplitude as a function of
the reduced temperature in
the first cell lie between two theoretical limits (Fig.\ref{fig_Thresh}.I):
curve {\bf A} represents the threshold in an infinite cell without the spectrum splitting
($r=1, \zeta=1/2$), and  curve {\bf B} shows the threshold in a finite cell
without reflection ($r=0$) and with the spectrum splitting ($\zeta=1/\sqrt2$).
An additional curve {\bf C} is plotted with the reflection coefficient
$r=0.7$ and the spectrum splitting
($\zeta=1/\sqrt2$).
It fits the data rather well for the most part of the experimental
temperature range.

For the temperature range close to $T_{\lambda}$, the experimental points
lie above the curve {\bf A}, that
presents the threshold in an infinite cell.
A reasonable explanation can be the following.
In the limit of $\Omega\eta^2/\gamma_2\ll 1$, the excited SS waves
have no spectra splitting and
propagate in the cell plane.
The  conditions of the SS generation by the FS standing waves vary
along the vertical axis.
Indeed, in the standing FS waves the regions of maximal amplitudes of
pressure and velocity are separated
in space.
It was shown in \cite{Murat-97}, that it is the pressure oscillations
in the FS waves that are
responsible for the SS generation.
So, the most preferable conditions for the SS wave generation are in the
antinodes of the pressure in the FS
standing wave.
Thus, the SS waves are excited in narrow layers near the FS
transducer surfaces and propagate along them.
At the same time, the SS bolometers are located in the first cell
near the central cell plane and are not sensitive to the SS waves
propagating near the transducer surfaces.
It means that the bolometers can only detect the SS waves well above
the onset, and the real threshold
value can not be measured.

The experimental data for the second cell agree with the theoretical
predictions rather well (Fig.\ref{fig_Thresh}.II).
All the experimental points lay between two curves {\bf B} and {\bf D}.
The curve {\bf D} presents the threshold for the cell with the zero
reflection coefficient, $r=0$, and the numerical
factor $\zeta=1/2$, that corresponds to no spectra splitting.
For small $\tau$, the experimental points are fitted by the curve
{\bf D}, and for large $\tau$ -by the curve
{\bf B}.
The crossover happens at the intermediate temperatures.

{\sl Splitting of SS waves spectra.}

The experiments in the first cell exhibit different types of the SS
spectrum in the different temperature ranges that
is well explained by the considerations of the finite size geometry
in the direction of the FS propagation and
by the discreteness of the $z$-component of the SS wave vector
(see Sect.\ref{sect_T_SpSplit}).
In the experimental temperature range, the attenuation of the SS
waves changes sharply, so the both limits
$\Omega\eta^2/\gamma_2\ll$  and $\gg 1$ can be observed.
Then indeed, if the $\Omega\eta^2/\gamma_2\ll 1$, that occurs closer
to $T_{\lambda}$, the discreteness
of the resonance states is smeared out.
This case corresponds to the symmetric decay with a single peak in
the SS spectra.
The opposite case $\Omega\eta^2/\gamma_2\gg 1$ leads to the discrete
resonance states and two peaks in the SS spectrum.
The temperature dependence of the frequency shift of the peaks
$\delta f=\pm (F/2)\eta^2$  is
plotted as dashed lines in Fig.\ref{fig_Spectr1} (here $F=\Omega/2\pi$).
It  describes the experimental data rather well.
Two peaks spectrum corresponds to the decay process in which one
of the SS waves propagates
normally to the FS wave direction, and both components of the FS
standing waves contribute to the parametric
excitation.  On the same plot, the temperature dependence of the SS
attenuation rate, $\gamma_2/4\pi$, is also shown
(Fig.\ref{fig_Spectr1}, dotted line).
For the temperature range, where two peaks are observed, the
following inequality is satisfied:
$\gamma_2/4\pi<\delta f$, while in the region of a single peak, the
relation is opposite:
$\gamma_2/4\pi>\delta f$.

The large difference in the measured peak amplitudes is probably due
to different SS modes corresponding
to each peak.
The peak with a lower frequency corresponds to the zero eigenmode along
vertical axis, that has a symmetric
distribution with respect to the central plane.
The peak with a higher frequency corresponds to the first mode, that is
antisymmetric.
The measured signal on the bolometer is a result of an interference.
The bolometer is symmetric with respect to the central plane.
So it can detect the antisymmetric mode only in the case when the
bolometer location deviates from the central plane symmetry.
Therefore, the higher frequency signal should have much lower amplitude.

{\sl Features of SS spectra in the second cell.}

The spectra of the SS waves in the second cell differ greatly from
the SS spectra in the first cell.
They are much more irregular.
These irregularities in the spectra do not change the temperature
dependence of the FS threshold amplitude
but do smear out the SS spectral splitting effect, observed in the first
cell.

There are two main factors that make two cells so different.
First, there is no reflection of the SS waves in the second cell in
contrast to the first one.
Second, the bolometers in the cells have very different angular
sensitivity as described in the Sect.\ref{sect_E_Cell1} and \ref{sect_E_Cell2}.

The striking result on the frequency shift of the SS spectrum peaks
just above the threshold of instability (Fig.\ref{fig_FrShift}) can be
explained
by anisotropy of the FS acoustic  field which had a non-zero component in the
horizontal plane.
The FS traveling waves could propagate in the direction shown by the
arrow in the inset (Fig.\ref{fig_FrShift}).
This could result from asymmetric FS reflection from the cell boundaries.
Both of the wave vectors of the FS standing wave components could be
slightly tilted.
In such a case, the SS spectrum measured on a bolometer, should be
shifted from exact $F/2$ value by
\begin{equation}
          \Delta f =   F (\frac {c_2}{c_1})\cos\alpha,
\label{eq_FrShift}
\end{equation}
where $\alpha$ is the angle between the FS wave vector in the horizontal
plane and the direction of the SS
wave propagation which is along the radial direction in the cell plane.
So, for the bolometers 2 and 8 one has $\cos\alpha < 0$, for the
bolometers 4 and 5 one has
$\cos\alpha>0$, and for the bolometer 7 one finds  $\cos\alpha\approx 0$.

As follows from Eq.(\ref{eq_FrShift}) the frequency shift due
externally broken isotropy in the cell plane is
proportional to $c_2/c_1$.
This dependence is different from the dependence due to spectral
splitting considered above,
where $\delta f\sim (c_2 /c_1)^2$.
Since the frequency shift due to anisotropy is larger than the
spectral splitting, that is probably the reason
why the latter effect was not observed in the second cell.

The experimental data on the frequency shift in the second cell were
fitted by Eq.(\ref{eq_FrShift}).
The dashed lines in Fig.\ref{fig_FrShift} show the result of the
fitting for each pair of the bolometers with a
single adjustable parameter $\alpha$.
The values of angles obtained from the fit for different bolometers, are the
following:
$\alpha_2 \approx \alpha_8 \approx(90+1.6)\deg$, $\alpha_4 \approx
(90-1.4)\deg$,
$\alpha_7\approx 90\deg.$

However, in spite of these nuisant effects it is still possible to
characterize nonlinear behavior of the
SS waves above the threshold in the second cell.

\subsection{\bf Nonlinear properties of SS waves above the threshold.}
\label{sect_R_Nonlin}

\subsubsection{\bf Nonlinear saturation mechanism.}
\label{sect_R_NonLinSatur}

As the FS amplitude increases above the threshold of the parametric
instability, the SS amplitude
starts to grow exponentially in time.
One of the central questions in this respect is a nature of a
mechanism of a SS amplitude saturation.
As we discussed in Sec. \ref{sect_T_NonlinSatur}, the theory
\cite{Murat-97} suggests that the 3W
resonance interaction is responsible for the saturation.
According to that theory, the SS wave intensity above the threshold
is defined by the nonlinear
attenuation and has the following dependence on the control parameter
of the instability:
$I_2\sim A/A_{th}-1$, see Eq.(\ref{eq_SSampl}).
Here $A$ and $A_{th}$ are the FS amplitude and its threshold value,
respectively.
This scaling is very different from that predicted and observed for
spin waves, where the  amplitude saturation occurs due to the dephasing
mechanism.
Then the spin wave intensity depends on the control parameter as
$I\sim\sqrt{(A/A_{th})^2-1}$  \cite{L'vov_b-94}.

The experiments in the first cell did not provide a clear cut answer
about the saturation mechanism.
An average intensity, particularly at large values of the control
parameter, $\epsilon=A/A_{th}-1$,
fluctuated so strongly with the control parameter, that it was
impossible to establish the intensity
dependence on $\epsilon $ (see Fig.\ref{fig_Intens}.I).
A possible mechanism for such strong fluctuations will be discussed below.

The data on the SS intensity as a function of $\epsilon$ in the
second cell can be analyzed quantitatively.
The typical plot of the SS intensity in a wide range of the FS
amplitude variation  is shown in Fig.\ref{fig_Intens2}.
This plot is undoubtfully exhibits linear dependence on the control
parameter
$I_2=g_{exp}\epsilon$,  where $g_{exp}$ is the parameter of the linear fit.
It was possible to measure this parameter in a wide temperature range
in the vicinity of $T_{\lambda}$
only for two bolometers, 4 and 8.
The dependences of $g_{exp}$ on the reduced temperature for these
bolometers are presented in Fig.\ref{fig_Slope}.
For the reduced temperature $\tau=3.3\times 10^{-4}$ one finds
$g_{exp} \approx 10^{-6} \mu W/cm^2=10^{-8}(W/m^2)$.

Thus, the data verified the theoretically found functional dependence
of the SS wave intensity above the threshold on the control parameter
presented by
Eq.(\ref{eq_SSampl}).
However, we found significant quantitative discrepancy both
in the  value of the coefficient in this expression
and its temperature dependence (see Sec.\ref{sect_T_NonlinSatur}).

In order to compare the experimental value of $g_{exp}$, we would
like, first, to make comments about a dimensionality of SS wave
packets in the first and the second cells. In both cells, the SS
waves propagate in the plane of the FS transducers. That defines
2D geometry of wave propagation. In the first cell dimensionality
of the wave pattern may be reduced to 1D, due to boundary
conditions. The resonance reflections from the bolometer and
heater substrates may select 1D pattern of the wave propagation.
 This
effect should be observed further away from $T_{\lambda}$. Closer
to $T_{\lambda}$, the dissipation length becomes very short, and
the role of the boundaries diminishes. In the second cell with the
non-reflecting boundaries the dimensionality of the wave packets
should be 2D. A nonlinear pattern selection may change it to 1D,
however, we do not see any evidence of such a phenomenon.

Even for the lowest theoretically predicted value of $g$ that corresponds to
1D
case, we found a large discrepancy with the experiment.
For the temperature $\tau=3.3\times 10^{-4}$, the experimental value
$g_{exp}$ is
450 times smaller than theoretically predicted one.
In 2D case the discrepancy even higher by the factor $\sqrt
{\omega_0/\gamma_0}$.
Comparison of the results in two cells does not improve the situation.
While the dimensionality of the first cell should be lower, and $g$ should be
smaller
correspondingly, one can see from the plots in Fig.\ref{fig_Intens}.I that
the
average slope of the SS intensity $vs$ $\epsilon$ in the first cell,
particularly
at small values of  $\tau$, is an order of magnitude larger  than in the
second
cell.
The inconsistency exists not only in the data for the different type
of the bolometers in two cells but  even for two different bolometers of the
same
type in the same cell (see data on two bolometers in Fig.\ref{fig_Slope}).
Moreover, as it is clear from Fig.\ref{fig_Slope}, a different
functional dependence of $g_{exp}$ on the reduced temperature was observed
for two
bolometers: for the  bolometer 4 one finds $g_4\sim \tau^3$, and for the
bolometer
8 one has $g_8\sim \tau^4$.
Both these power laws are very different from the
theoretically  predicted $g\sim \tau^{0.33}$
\cite{Murat-97}.

The reason for such great quantitative discrepancies between the theory and
the
experiment, and also inconsistencies of the results on the different bolometers
and
in two  cells may be explained by the following arguments.
There are two unknown factors that can influence the absolute value
of the SS amplitude measurements:
(i) an inhomogeneity of the bolometer sensitivity along  the fiber, and (ii)
a nonuniformity of a profile of a SS wave packet across the cell height.
The consequence of the latter factor on the measurements of the
instability threshold in the first cell has been already discussed above
(Sect.\ref{sect_R_Comp}).
A nonuniform spatial structure of a SS wave packet, that was not taken into
account by the theory\cite{Murat-97}, maybe another possible explanation of the
quantitative discrepancy.
For the sufficiently small $\tau$ the SS wave spectrum exhibits only one central
peak
(see Sect.\ref{sect_R_Comp}).
Then the SS waves propagate almost in the cell plane.
However, the SS wave packet has nonuniform amplitude distribution in  the
vertical
direction.
As was discussed above, the most preferable conditions for the SS wave
generation
are in narrow layers near the FS transducers, because the amplitude of the pressure
oscillations  there has a maximum.
Moreover, two SS waves, propagating along two transducer surfaces and
along different directions in the transducer plane maybe almost incoherent
and
produce a random  destructive interference on the bolometers.
All these factors can drastically reduce the measured SS intensity.

The measurements of the SS intensity require a bolometer calibration.
In our cell, it was possible to make only a static bolometer calibration,
namely
to measure the bolometer resistance {\sl vs} the cell temperature.
For experiments with the SS plane wave such calibration was sufficient to
get
correct numbers \cite{Goldner-93}.
However, in the experiment on the SS parametric generation the
distribution of the SS amplitude between the FS transducers is unknown.
If the SS wave packet profile across the cell is far from planar wave
and the SS  waves from different directions
are incoherent,  it is easy to get the measured signal ten and even
more times smaller than a real amplitude
of the SS wave for bolometers with inhomogeneous sensitivity distribution
along the fiber.
This gives correspondingly more than two orders of magnitude error in
the SS intensity measurements.
Then bolometers with different inhomogeneity can measure different SS
amplitudes for the wave packet.
The discrepancy between different bolometers can depend on temperature
because
both the bolometer sensitivity distribution and the SS wave amplitude
profile can
be a function of temperature.

Thus, in spite of the fact that the measurements of the SS intensity
above the instability threshold in
the second cell do not provide a possibility of quantitative
verification of the theoretical predictions
on the coefficient $g$ from Eq.(\ref{eq_SSampl}), these experiments
do provide information about the power
dependence of the SS intensity as a function of the control parameter.
In this way, the experimental results do verify the theoretical
prediction that the nonlinear attenuation is
the main mechanism of the SS amplitude saturation.
We can not exclude the possibility that the major mechanism of SS amplitude
satuartion is different from the nonlinear attenuation,  and the linear dependence of the SS
amplitude is the result of a coincidence of many other factors discussed above.
However, this is much less probable and we have no arguments to prove it.

\subsubsection{\bf Broadening of the SS spectrum above the threshold.}
\label{sect_R_SpBroad}

It was already pointed out, that the dynamic nonlinear behavior of
the parametrically generated SS above
the threshold is highly intermittent.
Such intermittent behavior differs greatly from a known example of
the parametrically excited surface
waves \cite{Cross-93} (see Sect.\ref{sect_T_Comp}).
In particular, the difference with pattern forming systems manifests
itself  in low frequency fluctuations of
the SS amplitude (Fig.\ref{fig_SSampl}).
The characteristic time of the fluctuations is much longer than the
relaxation time of SS waves,
$\gamma_2^{-1} $, and their traveling time, $L/c_{2}$. The behavior
similar to this was observed in some
cases of the parametrically excited spin waves \cite{L'vov_b-94}.

In the whole experimental temperature range, the SS frequency spectra
just above the threshold consist of
one or two instrumentally narrow peaks that broaden continuously as
the control parameter
increases.
In spite of the fact that the SS frequency spectra exhibit
qualitatively similar behavior in the entire
temperature range, it is possible to analyze them quantitatively only
in a narrow range of the
reduced temperatures between $2\times 10^{-4}$ and $3\times 10^{-4}$.
The range of the control parameter, however, is rather wide:
$0<\epsilon \lesssim 1.5$.
In this temperature range, the spectra have a single peak exactly at a
half the FS frequency, and the SS
attenuation length is less than the horizontal size of the cell.
The typical 3D plot of the SS spectral density as a function of the
FS amplitude and the frequency shift
$\delta f=f - F/2$ is shown in Fig.\ref{fig_Spect3d}.

The narrowness of the temperature range where good quality data are
observed, can be explained by several
reasons.
The ratio $l/L$ in this temperature range changes from $0.55$ to
$0.95$.
At $\tau<2\times 10^{-4}$
the SS attenuation length becomes
comparable with the distance between the FS
transducers ($d=3.9$ mm).
Therefore the SS waves that reach the bolometers, are generated near
the cell edge where a FS acoustic field
is inhomogeneous.
In this region, we observed qualitatively the same SS spectra shifted
a few Hertz to the right from the basic
SS frequency.

At $\tau>3\times 10^{-4}$ two other factors come into the play. The first
one is the resonance
cavity effects discussed in Sect.\ref{sect_T_SpSplit} and
\ref{sect_R_Lin}.
It causes the transition from one to two peaks spectrum. The second,
as the ratio $l /L$ exceeds unity, the lateral
boundary conditions become important.

At the threshold of the parametric instability, the SS wave spectrum
is very narrow and it starts widening as
the control parameter increases.
However, in the entire experimental range of the control parameter,
the SS spectrum remains very narrow
compared to other characteristic frequency scales.
The smallest among them is the SS attenuation rate, $\gamma_2 $, that is
about $40-60$ Hz.
The typical SS spectra for different values of the control parameters are
shown in Fig.\ref{fig_Trian}.
They have a finite width, $\Delta $, and exponentially decaying tails.
In semilogarithmic coordinates the spectra look like triangles with
equal sides.
The width of the spectrum grows as the control parameter increases.
In order to analyze the dependence of the spectral width on the
control parameter and temperature, we fit
the data by one of simple functions which possess the exponential
tails, namely
\begin{equation}
        N(f - F/2) = \cosh^{-1}\left( \frac{f- F/2}{\Delta} \right).
\end{equation}
 We emphasize here that the characteristic
width of the wave packet, $\Delta $, is
20-50 times smaller than $\gamma_2$ and is the
result of the nonlinear indirect 4W interactions
\cite{Murat-97}, as explained in the Sect.\ref{sect_T_broad}.

Fig.\ref{fig_SpWidth} presents the dependence of the spectral width
 as a function of the FS
amplitude for different temperatures.
At each temperature, the initial parts of the curves show a rather
weak dependence on the control
parameter.
Then in a wide range of the control parameter $0.1<\epsilon <1.5$,
the spectra width is fitted
rather well by a linear function of the FS amplitude:  $\Delta
=m(A/A_{0}-1)$.
$A_{0}$ differs from $A_{th}$ due to flat initial part of the curves,
as it is seen from the plot.
The temperature dependence of the slope $m$ is shown in the inset of
Fig.\ref{fig_SpWidth}.
It increases about twice in the narrow temperature range very similar
to the SS attenuation rate, $\gamma_2 $.
A comparison of the experimental, almost linear dependence of the
spectral width on the control parameter
with the theoretical predictions for different space dimensionalities
of the SS waves (see Eq.(\ref{eq_SpWidth})),
allows us to  conclude that the experimental situation is either 1D
or lies between 1D and 2D cases.
The temperature change of the prefactor in Eq.(\ref{eq_SpWidth}) is
also close to the experimentally observed
temperature dependence of the coefficient $m$.

\subsubsection{\bf Statistical analysis of the SS amplitudes.}

In the same temperature range, where the SS spectra
broadening was measured, we also conducted the statistical analysis
of the SS amplitude signals measured
by both bolometers (see Fig.\ref{fig_SSampl}).
We intended to answer the question whether the statistics of the SS
wave amplitudes was Gaussian.
The characteristic plots of the SS amplitude time series are shown in
Fig.\ref{fig_SSampl}.

One of the parameters that characterizes the deviation of the
amplitude distribution function from the
Gaussian one, is the ratio: $\Phi =M_{4}/M_{2}^{2}$, where $M_{2}$
and $M_{4}$ are the second and the
fourth order statistical moments:
$M_{2}=\left\langle \left( Z-\left\langle Z\right\rangle \right) \left(
Z-\left\langle Z\right\rangle \right) ^{\star }\right\rangle $ and
$M_{4}=\left\langle \left( Z-\left\langle Z\right\rangle \right) ^{2}\left(
Z-\left\langle Z\right\rangle \right) ^{\star 2}\right\rangle $, $Z$ is the
complex wave amplitude, $\left\langle \ldots \right\rangle $ means averaging
over time series.

A characteristic plot of $\Phi $ as a function of the FS amplitude at
a fixed temperature is shown in
Fig.\ref{fig_Stat}.
Below the threshold $\Phi = 2$, that provides an evidence of the Gaussian
distribution of complex amplitudes of
an experimental noise without any phase correlation.
Near and above the threshold $\Phi $ fluctuates strongly.
At $\epsilon >1/2$ it approaches an average value of about $3$.

The statistics is Gaussian if the fourth order correlation function
is equal to the sum of all possible products
of the pair correlation functions:
\begin{equation}
     \left\langle b_1b_2b_3^*b_4^*\right\rangle=\left\langle
b_1b_3^*\right\rangle
     \left\langle b_2b_4^*\right\rangle+\left\langle
b_1b_4^*\right\rangle\left\langle
     b_2b_3^*\right\rangle +\left\langle
b_1b_2\right\rangle\left\langle b_3^*b_4^*\right\rangle,
\label{eq_Stat}
\end{equation}
where $b_i=b({\bf k_1},\omega_1)$ is the SS wave amplitude for  given
${\bf k_i}$ and $\omega_i$.
Two first terms are products of the second order normal correlators:
\begin{equation}
\left\langle b_{{\bf k}_{1},\omega _{1}}b_{{\bf k}_{2},
\omega _{2}}^{*}\right\rangle =n_{{\bf k}_{1},\omega _{1}}\delta \left(
{\bf k}_{1}-{\bf k}_{2}\right) \delta \left( \omega _{1}-\omega _{2}\right),
\end{equation}
$n_{{\bf k}_{1},\omega _{1}}$ is the number of waves with given ${\bf
k_i}$ and $\omega_i$,
or the spectral wave density.
The third term is a product of the anomalous correlators:
\begin{equation}
\left\langle b_{{\bf k}_{1},\omega _{1}}b_{{\bf k}_{2},\omega
_{2}}\right\rangle =
\sigma _{{\bf k}_{1},\omega _{1}}\delta
\left( {\bf k}_{1}+{\bf k}_{2}\right) \delta \left( \omega
_{1}-\omega_{2}\right),
\end{equation}
The anomalous correlator differs from zero for parametrically
generated waves because
a pumping field produces pairs of waves, and the waves inside one
pair are strongly correlated.
It was shown theoretically \cite{L'vov_b-94}, that in a steady state
$\left| \sigma _{{\bf k},\omega }\right| =\left| n_{{\bf k},\omega
}\right| $, and that leads
to the following relation between statistical moments:
\begin{equation}
M_{4}=3M_{2}^{2}.
\end{equation}
The relation between the fourth and second moments with zero
anomalous correlator is
different: $M_{4}=2M_{2}^{2}$.
Below the threshold of the parametric instability the measured signal
is the electronic noise.
This noise certainly has no anomalous correlator, and it is Gaussian,
i.e., $\Phi=2$.
Close to the threshold the value of $\Phi$ fluctuates strongly.
One of the reasons for such fluctuations is not sufficient time of the
statistical measurements.
The characteristic time scale of the amplitude fluctuations for small
$\epsilon$ is comparable with the
measuring time (see Fig.\ref{fig_SSampl}).
For large $\epsilon$, $\Phi$ becomes  equal to $3$, and this is a
crucial evidence of the Gaussian statistics
of the SS wave amplitudes.
This transition in the value of $\Phi $ from $2$ below the threshold
to $3$ above it, indicates a very crucial
manifestation of a physical mechanism for the parametric instability,
namely, appearance of a parametrically
initiated wave pairing with a strong phase correlation inside of each
wave pair.

We verified the Gaussian statistics only for the
correlators with a time delay and a space shift both
equal zero.
So we got necessary but not a sufficient proof of the Gaussian statistics.
However, it is the first experimental evidence of the validity of the
theoretical hypothesis about the Gaussian
statistics for the parametrically generated waves.

The Gaussian statistics of the SS amplitude fluctuations above the
threshold verified in the experiment,
unambiguously points at the kinetic theory to describe the behavior
of the parametrically excited SS waves and
provides a foundation for its applicability.
In this approach, one introduces pair correlation functions of
parametrically excited waves and, using
a hypothesis of the Gaussian statistics of wave amplitudes, splits
higher order correlation functions, that
appear in nonlinear terms, into products of the pair correlation
functions \cite{L'vov_b-94}.
The kinetic theory also predicts that a spectral packet has
universal exponential tails and its width is small
due to relatively weak 4W scattering processes.
Therefore, the observed shape and width of the packet confirm the
main results of the kinetic theory.

We also measured  the crosscorrelation function of the amplitude signals,
$Z_1$ and $Z_2$, from
two bolometers in the first cell:
\begin
{equation}
C(\theta)=\frac{\left\langle \left( Z_1(t)-\left\langle
Z_1\right\rangle \right) \left(
Z_2(t-\theta)-\left\langle Z_2\right\rangle \right) ^{\star }\right\rangle}
{\sqrt{\left\langle  |Z_1(t)-\left\langle Z_1\right\rangle|^2
\right\rangle\cdot \left\langle
|Z_2(t)-\left\langle Z_2\right\rangle|^2 \right\rangle }}.
\end{equation}

The bolometers are sensitive to the waves propagating in almost
perpendicular directions.
The absolute value of the normalized crosscorrelation function is found
to be  $|C(\theta)|<0.1$ for all delays
$\theta $.
That can be explained by an absence of angular correlations between the SS
waves, and once more
indicates a random phase distribution of  wave amplitudes already
discussed above.
Moreover, we were able to measure  correlations of the SS signals
obtained by the different
bolometers also in the second cell.
We found no significant correlations between the signals on the
bolometers near the threshold instability.
That indicates no tendency to pattern formation.
This observation agrees with the theoretical conclusion about the
absence of a long-range order at an angle
of resonantly interacting SS waves.
As we suggested above (see Sect.\ref{sect_T_NonlinSatur}),
this state, particularly very close to the instability threshold, can be described as a turbulent
crystal \cite{Newell-93}.

The 4W resonance interaction is present in any medium with
parametrically driven waves.
Thus the kinetic theory predicts the universal spectral broadening in
such systems irrespective of the wave
dispersion law.
We do not exclude a possibility that this universal spectral
broadening can be observed in the capillary waves
too.
This would be possible if the resolution in frequency measurements is
drastically improved compared to that
reached in previous experiments, to be much better than the ratio
$\gamma/\omega$, which for the
Faraday ripples is of the order $10^{-2}$.

\subsection{\bf Experiment on simultaneous FS and SS pumping}
\label{sect_R_SSpump}

\subsubsection{\bf Acoustic phase conjugation in superfluid helium.}

The experiments in the second cell with a simultaneous FS and SS
pumping reveal a new effect in the
dynamics of the SS waves below the parametric instability threshold.
An analog of the well-known in optics phenomenon of a phase conjugated (PC)
mirror was observed in the superfluid
helium for the SS waves.
A mirror built on this effect, in a contrast to a conventional one,
reflects an incident wave in such a way that
a reflected wave is always directed opposite to an incident one.

One of the differences between the first and the second cells
essential for such experiments
is the presence of the bolometer located between the SS emitter and the
cell.
Thus, in the second cell, we had an opportunity to measure the SS
wave amplitude emitted from the heater
before it entered into the cell.
On the same bolometer we could measure an amplitude of the SS wave
that was coming back from the cell.
Another great advantage of the second cell, which allows to perform
the experiments on the phase
conjugation, is a wide angle sensitivity of the fiber bolometers and
absence of a reflection for the SS waves.

{\sl Theoretical background.}

A PC mirror has several unique properties compared to an ordinary mirror.
It reflects an incident wave back for any incident angle.
The conjugated wave can have a larger amplitude than the incident one.
But it is this time-reversed phase property of the reflected wave
that makes the optical PC so potentially useful
for a host of interesting applications, and particularly for correction
of wavefront distortions \cite{Fisher_b-83}.
A common but not a sole realization of PC in optics is the
mirror, based on 4W interactions \cite{Fisher_b-83}.
It can also be realized through 3W interactions.
In the latter case, PC has been observed in various wave systems
manifesting sufficiently strong nonlinear
interactions, e.g. microwaves \cite{Shih-90} and acoustic waves
\cite{Thompson-71}.
In fact, the first observation of PC in acoustics was made long
before the observation of the optical PC
\cite{Thompson-71}.
PC in acoustics results from the interaction between sound waves and
the various types of collective
oscillations in solids.
The interaction of a sound wave (either longitudinal or shear) with
electromagnetic waves in piezoelectrics or
magnets, the phonon-plasmon interaction in piezoelectric
semiconductors, and the interaction of
electromagnetic waves with spin waves in magnets are a few of many examples.
At large enough amplitudes of an external field, these systems
exhibit a space-homogeneous parametric
instability, as was already described above.

The PC phenomenon of the SS waves in the superfluid helium can be
observed below as well as above the
parametric instability threshold.

An incident SS wave with a half the frequency of the FS wave can be
amplified by the FS pumping wave, generating a PC
wave in the opposite direction.
Indeed, as can be easily seen from the conservation laws of
Eq.(\ref{eq_ParamDec}), at
${\bf |K|\ll |k_{1,2}|}$ one gets  $\omega_{1,2}\approx \Omega/2$,
and the parametrically generated SS waves
propagate in almost  opposite directions ${\bf k_2\simeq -k_1}$ with the
conjugated phases.

Two main factors distinguish our system from the common manifestation
of PC in optics.
First, PC in optics is usually examined only below the onset of
spontaneous oscillations (instability), which is
unattainable at the currently available laser intensity \cite{Fisher_b-83}.
Second, the optical systems are of a very low dissipation, so that $l/L\gg 1$.
Our system is in the range of $l/L\geq 1$ depending on the temperature,
so the dissipation is of a crucial
importance for the conjugated wave generation.

Two linear (in respect to the incident and conjugated waves
amplitudes) problems can be formulated in
regards to the parametric instability of FS:
1) determination of the threshold of a spontaneous SS wave generation
via a 3W interaction process,
2) generation of the conjugated SS wave below the instability
onset due to the nonlinear interaction
of the incident SS wave with the FS pumping field.

The first problem, discussed in Sect.\ref{sect_T_ParInst} and
\ref{sect_R_Lin}, deals with the spatially
uniform, rotationally  invariant state in a FS resonance cavity.

The second problem is related to a system with its rotational
invariance externally broken by the incident SS
wave having a spatially dependent amplitude.
The relevant question here is: what is the amplitude value of the
conjugated SS wave which is generated as a
result of the 3W interaction?
Depending on the boundary conditions and the value $l/L$ for the same
cell geometry this problem can be
described in two ways.
First method is similar to the PC paradigm in optics \cite{Fisher_b-83}.
One can consider the inhomogeneous problem for a space-dependent
amplitude distribution of the probe and the
conjugated waves in a resonance cavity with a dissipation.
In this case the nonlinear process of the PC wave generation via the 3W
interactions (with non-zero wave number mismatch
$\Delta k=\delta/c_2$) is described by the following set of linear
steady state equations
\begin{eqnarray}
       (\partial/\partial x+\gamma_2/c_2) b_1 + i\kappa b_2^* \exp
(i\Delta k x)=0, \\
       (\partial/\partial x+\gamma_2/c_2) b_2 + i\kappa b_1^*\exp
(i\Delta k x)=0,
\end{eqnarray}
and the boundary conditions for both incident and conjugated waves:
\begin{equation}
      b_1(0)=b_0, \hspace{0.5in} b_2(L)=0.
\end{equation}
Here $\kappa\equiv aU/c_2$, $b_{1,2}$ are the incident and conjugated
SS wave amplitudes, respectively,
$0\leq x \leq L$ is the coordinate in the direction of the SS wave
path, and  $b_{1,2}(0)$ are the
amplitudes of the incident and the reflected SS waves at the cell
entrance, respectively.
Rather straightforward calculations lead to the following expression
for the nonlinear power reflection
coefficient, defined as $r=|b_2(0)|^2/|b_1(0)|^2$:
\begin{equation}
    r=4\kappa^2\frac{\cosh(2gL)-\cos(2fL)}{D},
\label{eq_PCrefl1}
\end{equation}
where
\begin{eqnarray}
D & = &\exp(-2gL)\left[(f-\Delta)^2+(g-\gamma_2/c_2)^2\right] + \nonumber \\
    &+&\exp(2gL)\left[(f+\Delta)^2+(g+\gamma_2/c_2)^2\right] + \nonumber\\
    &+&4(g^2-(\gamma_2/c_2)^2)\cos(2fL)
    +4(f\gamma_2/c_2-g\Delta)\sin(2fL),  \label{eq_PCrefl2}\\
      f^2-g^2 &=&\kappa^2+\Delta^2-(\gamma_2/c_2)^2, \hspace{0.5cm}
    fg=\Delta\gamma_2/c_2.  \label{eq_PCrefl3}
\end{eqnarray}
Eqs.(\ref{eq_PCrefl1}-\ref{eq_PCrefl3}) are transformed to the known
expression for the power reflection coefficient in the
optical PC via a nearly-degenerate 4W mixing at $\gamma\rightarrow 0$
\cite{Fisher_b-83}.
At this limit, in the region $\pi/4<\kappa L<\pi/2$, the intensity of
the PC wave exceeds that of the incident one, i.e.,
it is the regime of the PC coherent amplifying reflector.
The dissipation cuts down the amplification and shifts the  amplification
region closer to the onset of the parametric instability.
Thus both the dissipation and the frequency mismatch result in a decrease
of the PC reflectivity, and the conjugator behaves as a
narrow bandpass acoustic filter.
In the limit $a/a_{th}\ll 1$,
Eq.(\ref{eq_PCrefl1}-\ref{eq_PCrefl3}) can be simplified
\begin{equation}
   r =
4\kappa^2\frac{\cosh(2gL)-\cos(2fL)}{\exp(2L\gamma_2/c_2)\left[(\delta/c_2)^
2+(\gamma_2/c_2)^2\right]}.
\label{eq_PCrefl4}
\end{equation}
This limit can be always achieved, and the expression
(\ref{eq_PCrefl4}) can be easily  analyzed.
Indeed, it follows that:
(i) a reflected wave amplitude, $|b_2(0)|$, is proportional to a FS
amplitude, $a$;
(ii) $|b_2(0)|$ is proportional to an incident wave amplitude, $|b_1(0)|$;
(iii) a reflection coefficient has a resonant dependence on frequency
mismatch, $\delta$.
The resonance curve depends on the SS attenuation rate, $\gamma_2$, and
the cell size, $L$.

However, as our calculations and comparison with the experiments
show, the expressions either
of Eqs.(\ref{eq_PCrefl1}-\ref{eq_PCrefl3}) or Eq.(\ref{eq_PCrefl4})
do not fit the experimental data.
This is, probably, due to the relatively large dissipation and,
particularly, due to the nonreflecting lateral boundaries.

The second method is to consider a homogeneous problem for the probe
and the conjugate waves with a linear decay.
Thus, in this case the nonlinear process of the PC wave generation via the 3W
interaction  is described by the following set of linear
equations:
\begin{eqnarray}
\left[\partial/\partial t +\gamma_2 +i\omega_1\right]b_1 +
iUab_2^*=0, \label{eq_PChom1}\\
\left[\partial /\partial t
+\gamma_2-i\omega_2\right]b_2^*-iU^*a^*b_1=0  \label{eq_PChom2}
\end{eqnarray}
Here, the incident and the conjugated SS waves have the amplitudes $b_1$ and
$b_2$ and the frequencies $\omega_1$ and
$\omega_2$, respectively, and
$\omega_{1,2}=\Omega/2 \pm \delta$, where $\delta$ is the frequency
shift of the parametrically generated SS waves.
The nonlinear power reflection coefficient is obtained from
eqs.(\ref{eq_PChom1}-\ref{eq_PChom2}) as:
\begin{equation}
     r = \frac {|Ua|^2}{\gamma_2^2 + \delta^2}.
\label{eq_PCreflcoef1}
\label{eq_PChomrefl}
\end{equation}
Here $b_{1,2}(0)$ are the amplitudes of the incident and the reflected SS
waves, respectively, at the cell entrance.

In order to describe the experimental data by Eq.(\ref{eq_PChomrefl}),
obtained in a finite lateral geometry FS resonance cavity,
one can rewrite Eq.(\ref{eq_PChomrefl}) by incorporating
Eq.(\ref{eq_Ath}) in the following form:
\begin{equation}
       r=(\frac{a}{a_{th}})^2 \frac{\zeta^2[1+(\xi
l/L)^2]}{1+(\delta/\gamma_2)^2}.
\label{eq_PCreflcoef2}
\end{equation}
This approach is justified in the linear regime below the parametric
instability threshold.
The effective attenuation $\Gamma=\gamma_2-\sqrt{|Ua|^2-\delta^2}$
tends to zero at the threshold and
becomes negative above it, causing the SS wave amplitude to diverge
exponentially.
Saturation  occurs due to higher-order nonlinear effects of the SS wave
interaction (see
Sect.\ref{sect_T_NonlinSatur} and \ref{sect_R_Nonlin}).

The main theoretical predictions for the PC  waves generated
parametrically {\sl below} the onset are:\\
1. The amplitude of the conjugated SS wave is proportional to: (a)
the FS amplitude at fixed value of the frequency shift, and
(b) the amplitude of the incident SS wave.\\
2. The sum of the phases of one FS and two SS waves involved in the
resonance interaction, is $\phi+\phi_1+\phi_2=\pi/2$.
For a pair of the SS waves this leads to  $\phi_1+\phi_2=const.$\\
3. The power reflection coefficient as a function of the frequency
shift has a Lorentzian shape with a width equal to the SS
wave linear attenuation rate.

{\sl Results and discussion.}

As in the previous experiments, we performed the SS spectrum measurements
in a narrow bandwidth around
$\Omega/2$ using the lock-in amplifier fixed at this reference
frequency (see Sect.\ref{sect_E_SpMeas}).
The experiment was conducted in the second cell at a fixed
temperature by pumping the acoustic cavity with the FS
waves at the frequency $F$ and the amplitude below the threshold value for
the parametric instability.
At the same time, a small amplitude SS wave was emitted by the heater.
In order to separate SS signals coming directly from the heater and
those coming from the cell, the incident wave
frequency was shifted from  $F/2$ to $f_1=F/2+\delta f$.

A bolometer B8, located far from the heater (see Fig.\ref{fig_Cell2}),
detected only the SS waves with the frequency $f_1$.
A bolometer B4, on the other hand, located  between the heater and the
cell, detected two signals: one with the frequency $f_1$ coming from the
heater,
and another with the frequency $f_2=F/2-\delta f$ coming from the
cell.
This dual detection is clearly seen from the power spectrum of
the signal read by the bolometer B4 (see inset of Fig.\ref{fig_PCMampl}).
The spectra were measured at the reduced temperature $\tau=7.07
\times 10^{-4}$ and the FS amplitude $A=75$ Pa.
This amplitude was below the threshold value $A_{th}=162$ Pa, measured
at the same temperature.
The dependence of the amplitudes of both peaks at the same temperature
on the FS amplitude is presented in Fig.\ref{fig_PCMampl}, the frequency
shift is
$\delta f=-12$ Hz.
The amplitude of the incident SS wave $b_1$, arriving directly
from  the heater on the bolometer B4, clearly does not depend on the FS
amplitude,
since these waves do not interact in  the region between the heater and the
resonance cavity.
On the other hand, the amplitude of the SS waves arriving from the
cell, $b_2$, is a linear function of the FS  amplitude below the parametric
instability threshold, as follows  from the theory, (see
Eq.(\ref{eq_PCreflcoef1})),  and is also a linear function of the SS wave
amplitude emitted by the heater $b_1$.
The estimate of the slope of $b_2/b_1$ as a function of $A/A_{th}$
from Fig.\ref{fig_PCMampl} gives $1.7 \pm 0.15$.
This value is in a fair agreement with the theoretical value of 2.05
obtained from Eq.(\ref{eq_PCreflcoef1}).
Thus, the first property of the conjugated SS signal, following from
Eq.(\ref{eq_PCreflcoef1}), is quantitatively verified.

Particular efforts were made to verify the phase relation between the
incident and the conjugated SS waves.
The phases of each of the SS signals at the frequencies $f_1$ and $f_2$
were arbitrary with respect to the pumping
field, and changed randomly from one measurement set to another (open
circles in Fig.\ref{fig_PCMphase}).
However, the sum of the two phases of both signals was constant for
all FS and SS incident wave amplitudes (Fig.\ref{fig_PCMphase}).
The standard deviation from the average value of the phase sum was
$\delta(\phi_1 +\phi_2)=0.15$ rad, as compared to
the uniform distribution of the phase,  in a $2\pi$ bandwidth, for
each separate signal.
This result is the main evidence for the PC in the SS waves.

The dependence of the nonlinear reflection coefficient, $r$, on the
frequency shift, $\delta f$, at the reduced
temperature $\tau=7.07 \times 10^{-4}$ and the FS amplitude $A=75$ Pa
is presented in Fig.\ref{fig_PCMfreq}.
Fitting the experimental data by the Lorentzian function gives a
bandwidth of $\Delta=14$ Hz compared with the
theoretical width $\Delta_{th}=19$ Hz.
The theoretical width is defined solely by the SS attenuation in an
infinite cell at the experimental values of
$\tau$ and the SS frequency $F/2\approx 20$ kHz.
The main reason for the discrepancy is a near-field configuration,
used in the experiment.
The latter can lead to distortions of the linear decay rate.

In conclusion, we have presented the first direct evidence of PC
in the SS waves, as a result of the 3W-interaction between the FS pumping field and
the SS
waves,  below the threshold of the parametric instability.
Since PC is a fairly general phenomenon in parametrically generated
waves, we are convinced that it can be observed
in other parametrically driven systems.
One of these easily-accessible systems is that of the surface waves
parametrically excited by a vertical vibration
\cite{Cross-93}.
An obvious advantage of the latter system is an easy visualization
of the surface waves which can be
used to verify the PC effect experimentally.

\subsubsection{\bf Experiments on simultaneous FS and SS pumping
above the threshold.}

One of the powerful tools to probe a state of many waves is to
measure its response on small perturbations, namely on an externally
excited,
small amplitude wave. An independent SS wave
emitted from a heater, propagates through the nonlinear
medium of the interacting waves, and interacts with them.
We study experimentally the response of the parametrically generated
SS waves above the threshold on a small
perturbation in both cells.

{\sl Amplification of externally generated SS waves by FS above the
threshold.}

The first experiment on an amplification of the SS waves by FS was conducted
in the first cell.
The small amplitude SS wave was emitted from the plane heater (see
Fig.\ref{fig_Cell1}) at exactly a half the FS
frequency, $F/2$.
The signal was measured on the bolometer located on the opposite side
of the cell.
The SS intensity was measured as a function of the FS amplitude for the
different initial SS amplitudes and the different reduced
temperatures, $\tau$.

The amplitude of the emitted wave was calculated by using the data for the SS
wave attenuation \cite{Mehrotra-84} and the
measured SS wave amplitude on the bolometer.
The typical plots of the  SS intensity, $I$, as a function of the FS
amplitude, $A$, for different values of a heat flux from the heater,
i.e., the different SS amplitudes of the probe wave, are shown at
$\tau=3.32\times 10^{-4}$ in Fig.\ref{fig_SubPump}.
Each plot can be approximated by two linear regions: the first region
for sufficiently small FS amplitudes below the instability
onset, and second one is above the threshold.

For sufficiently small FS amplitudes, the linear attenuation and the 3W
interactions between the FS and SS waves
define the SS amplitude.
Below the threshold of the parametric excitation the effective attenuation rate
is
\begin{equation}
   \Gamma=\gamma_2-aU
\end{equation}
It can be rewritten as $\Gamma=\gamma_2(1-a/a^{\infty}_{th})$.
Then the SS wave intensity on the bolometer far below the threshold
can be expressed as
\begin{equation}
   I = I'_0(\frac{A}{A_{th}}+1), \hspace{0.5cm}
I'_0=I_0\exp(-\frac{2L\gamma_2}{c_2}),
\label{eq_SubpInit}
\end{equation}
where $I'_0$ is the SS wave intensity, reaching the bolometer without
FS pumping.
Using Eq.(\ref{eq_SubpInit}) to fit the initial part of plots in
Fig.\ref{fig_SubPump}, one gets the threshold FS
amplitudes and the SS wave intensity on the heater, $I_0$.

The SS wave amplitude above the threshold is defined by the nonlinear
attenuation,  discussed
in Sect.\ref{sect_T_NonlinSatur} and \ref{sect_R_Nonlin}.
Similar to the results on the SS intensity, discussed there, the SS
intensity slope, $g_{exp}$, was found to be
much smaller than the theoretically predicted one \cite{Murat-97}.

However, a new unexpected giant amplification of the SS probing waves
by the FS pumping was observed in the first
cell.
The SS intensity as a function of the control parameter, $\epsilon$,
on the plots with different $I_0$  and for
different $\tau$ was fitted by $I_2=g_{exp}(I_0,\tau)\epsilon$.
The threshold, $A_{th}$, was found to be almost independent on the
initial SS intensity, $I_0$, and was
the same as measured in the previous experiments without a simultaneous
SS pumping (see Sect.\ref{sect_R_Lin}).
As it was discussed before, the SS intensity above the threshold
fluctuates strongly in the first cell without a
simultaneous SS pumping (see Sect.\ref{sect_R_SpBroad}).
With the simultaneous SS pumping at a sufficiently small SS
amplitude, the intensity of the measured signal still
remains fluctuating.
As the SS pumping amplitude increases the fluctuations are
suppressed, and the slope $g_{exp}$
strongly increases.
Although, the maximum value of the slope is still remained lower than
the theoretically
predicted value \cite{Murat-97}. Plots of the dependence of the SS
intensity slope as a function of the SS probe
wave intensity for the different reduced temperatures are presented in
Fig.\ref{fig_SubpSl}.

Very close to $T_{\lambda}$, the SS intensity slope strongly depends
on the initial SS pumping intensity.
The value of $g_{exp}$ changes few orders of magnitude as $I_0$ increases.
This effect becomes weaker as $\tau$ increases.
The data $g_{exp}(I_0)$ for various $\tau$ is plotted in the
logarithmic scale in Fig.\ref{fig_SubpSl}.
These experimental data can be fitted by a function $g\sim
I_0^{\nu}$, where the index $\nu$ in its turn is a function
of the reduced temperature.
The inset in the Fig.\ref{fig_SubpSl} shows the dependence $\nu(\tau)$.
This dependence can be fitted by $\nu=\tau_0/\tau-\nu_0$.
The fit parameters were found to be : $\tau_0=5.2\times 10^{-4}$ and
$\nu_0=1$.
So, finally one has $\nu=\tau_0/\tau-1$.
This data representation has no theoretical background whatsoever.
However, it allows to make some speculations about possible reasons
for the giant amplification of the SS probe waves.

As we already discussed above (see Sect.\ref{sect_R_NonLinSatur}),
one of the possible explanations of huge
quantitative  discrepancy between the theoretically predicted value for $g$
and the experimental ones for $g_{exp}$,
measured without the  SS simultaneous pumping, may be the nonuniform
spatial structure of a SS wave packet that was
not taken  into account by the theory \cite{Murat-97}.
For the sufficiently small $\tau$, the SS wave spectra exhibit only one
central peak (see Sect.\ref{sect_R_Lin}).
Then the SS waves propagate almost in the cell plane.
However, a SS wave packet has nonuniform amplitude distribution in
the vertical direction.
As was discussed above, the most preferable conditions for the SS
wave generation are in narrow layers near the FS transducers, because the
amplitude
of the pressure oscillations has a maximum there.
On the other hand, the SS bolometers in the first cell are located in
the cell midplane and thus have low
sensitivity to the waves propagating close the FS transducer surfaces.
Moreover, two SS waves, propagating along two transducer surfaces
maybe almost incoherent and have a random
destructive interference on the bolometers.
All these factors could drastically reduce the measured SS intensity,
as we already discussed in
Sect.\ref{sect_R_NonLinSatur}.

In the experiment with the simultaneous FS and SS pumping the plane SS
wave, emitted by the heater, had a uniform
amplitude distribution in the vertical direction.
Such a wave interacted with the FS wave field and could lead to a
redistribution of a SS packet intensity at the
bolometer.
The emitted SS wave can also synchronize phases of two packets
propagating along two FS transducers
and suppress  the fluctuations of interference signal.
These factors alone can increase greatly the SS wave intensity
measured on the bolometer.
This effect should be certainly sensitive to initial SS wave
distribution and, thus, to the reduced temperature.

At large $\tau$, the SS wave spectra became splitted, and one wave in
a pair of the SS waves propagates out of
the  cell plane.
So the emitted SS wave from the heater may not have such strong
influence on the SS wave intensity
redistribution.
The reduced temperature at which the influence of the SS pumping wave
on the intensity slope $g_{exp}$
vanishes ($\tau_0=5.2\times 10^{-4}$, found from the fit of the data in
Fig.\ref{fig_SubpSl} ) lies surprisingly close to the value of $\tau$, at
which
the transition  from one peak to the splitted type of SS spectrum occurs.
The SS reflection from the lateral boundaries becomes important, and discrete
levels  come into the play.
At $\tau>\tau_0$ the slope $g_{exp}$ does not change with $I_0$ variations.
In this temperature range, another effect which modifies the SS  intensity
measurements, takes over.

{\sl Nonlinear SS wave resonances in the first cell with lateral
reflection.}

A typical plot of the SS wave intensity as a function of the FS amplitude
for a given SS pumping amplitude at
$\tau=5.21\times 10^{-4}$ is shown in Fig.\ref{fig_SubpInt}.
As seen clearly from the plot, the SS intensity strongly oscillates
as the FS amplitude changes.
The most surprising fact is that for some values of the control
parameter the SS wave intensity reaches
zero values.
One of the reasonable explanations can be the effect of nonlinear
renormalization of the SS wave number due to
the 3W interactions \cite{L'vov_b-94}.
In all nonlinear wave systems, the 3W interactions renormalize the wave
frequency.
It means that the dispersion relation becomes dependent on the
amplitude of the parametrically excited waves as
\begin{equation}
     \omega_{nl}=\omega(k)+2\int T(k,k') b(k')^2 dk',
\label{eq_NonlinFrShift}
\end{equation}
where $\omega(k)$ is the linear dispersion relation, $T$ is the 3W
interaction matrix element, and $b(k')$
is the SS wave amplitude.
In the case of the parametric wave generation, the pumping
frequency, $\Omega$, is fixed, and frequency of the
parametrically excited waves always satisfies the condition
$\omega_{nl}=\Omega/2$.
Then the Eq.(\ref{eq_NonlinFrShift}) leads to the renormalization  of the
wave vectors of the excited waves to satisfy the frequency
relation.
In the case of a cell opened in a lateral direction, such an effect
does not influence on an average intensity of parametrically
excited waves.
However, in a resonance cavity it leads obviously to resonance
oscillations of the wave intensity, detected
by a bolometer.
In the first cell at the reduced temperature larger than $\tau\approx
5\times 10^{-4}$, the
finite size effect became significant since the SS dissipation length
became larger than the cell size,
$l/L>1$ (see Sect.\ref{sect_R_Comp}).
Then the SS reflection at the lateral boundaries could modify the
dependence of the SS intensity on the FS amplitude.
Thus, the probable explanation of the strong SS intensity oscillations as
a function of the FS amplitude could be the
nonlinear renormalization of the dispersion relation of the
parametrically generated SS waves and influence of the
reflecting lateral boundaries on the SS measured intensity.
We would like to emphasize here that a similar effect of the strong SS wave
intensity oscillations but less pronounced,
was  observed in the first cell even without an additional SS pumping,
as we already pointed out in the
Sect.\ref{sect_R_Lin} and Fig.\ref{fig_Intens}.

Let's estimate the value of the control parameter variation,
$\Delta\epsilon$, which is necessary to switch
from one to another, close-by resonance SS modes.
In the case of the SS parametrically driven waves,
Eq.(\ref{eq_NonlinFrShift}) can be written as \cite{Murat-97}
\begin{equation}
    \omega_{nl}=\omega+Re\Sigma,
\end{equation}
where $Re\Sigma$ is the real part of the correction for the
self-energy function $\Sigma$, calculated in
Ref.\cite{Murat-97}.
For one-dimensional geometry the calculations give
\begin{equation}
  \Delta\omega\equiv\omega_{nl}-\omega=\frac{4B^2b^2}{\omega}\ln(\frac{\omega^
2}{320\gamma_2^2}),
\end{equation}
where $b^2=I_2/c_2\omega$ and $B$ is defined in Eqs.(20-22).

By substituting $I_2$ and $g$ from Eqs.(\ref{eq_SSampl} and \ref{eq_SSAmplSlope})
one gets
\begin{equation}
\frac{\Delta\omega}{\omega}=3.44 (\frac{\gamma_2}{\omega})^2\Delta\epsilon.
\end{equation}
This nonlinear frequency shift causes a change in  the nonlinear wave length:
$\frac{\Delta\omega}{\omega}
=\frac{\Delta\lambda}{\lambda}=(L/\lambda+1)^{-1}$.
Then the variation in the control parameter necessary for switching
to a neighboring SS mode is
\begin{equation}
   \Delta\epsilon\approx
\frac{1}{3.44}\frac{\lambda}{L}(\frac{\omega}{\gamma_2})^2\approx
50\gg 1.
\end{equation}
Thus again, we obtain large quantitative discrepancy between the
theoretical  estimates, based on the 3W interactions, and the experimental
results. It
is feasible that  it has the same source as the discrepancy in the value of
the
nonlinear coefficient $g$  for the amplitude saturation, which also results
from
the 3W interactions.

Another possible mechanism to explain the phenomenon is a
redistribution of the SS intensity between different  azimuthal modes due
to increase of a nonlinear SS
interaction as the FS pumping amplitude raises.  The azimuthal Bessel
modes of the SS waves have much smaller
differences between higher order modes in the spectrum  than the
longitudinal modes.
Then one can expect that the energy redistribution between the modes due
to the nonlinear interactions can bring into the resonance
different modes for different values of the FS wave amplitude.
Unfortunately, one cannot quantitatively estimate this effect.

\section{\bf Conclusions.}
\label{sect_Concl}

We present the results of the experiments on the parametric generation
of the SS wave by FS in the superfluid helium
in two resonant cavities with the different lateral boundary conditions
and the different angular sensitivity of the SS detectors.
There are three main subjects which were studied:\\
(i) the temperature dependence of the parametric instability threshold
and the SS spectra at the onset;\\
(ii) the mechanism of the wave amplitude saturation and the nonlinear
properties of the SS waves above the
threshold, their statistical and spectral characteristics;\\
(iii) the interaction between the independently pumped SS wave and
FS: SS phase conjugation below the
threshold, and the giant amplification of the parametrically generated SS
waves above the threshold.

Comparison of the presented experimental results with the theory,
reveals the following.
The temperature dependence of the FS threshold amplitude for the
onset of the parametric instability in
a resonance cavity of a finite lateral size agrees well with the
theoretical predictions (see Fig.\ref{fig_Thresh}) without any fitting
parameter.
The theory quantitatively explains also the experimentally observed
transition from a single line SS wave
spectrum close to $T_{\lambda}$, to a spectral spitting further from
$T_{\lambda}$.
It results from lifting of a degeneracy in the SS wave generation
process when the SS dissipation rate becomes
smaller than the frequency difference between two discrete resonance
modes in the SS wave spectrum in the
resonance cavity.

In regards to the nonlinear properties of the SS wave ensemble,
quantitative comparison is less successful.
The strong fluctuations of the SS amplitude in the first cell, probably
due to the lateral boundary reflections
do not permit to get definite conclusions about its functional dependence on
the FS pumping amplitude.
In contrast, the results from the second cell provide an evidence of the
mechanism of the SS amplitude saturation
based on the functional dependence of the SS wave intensity on the control
parameter for the different reduced
temperatures.
The linear dependence of the SS wave intensity in a wide range of the
control parameter points out on the
nonlinear attenuation due to the 3W resonance interactions of the SS waves as
a main mechanism of the SS amplitude
saturation in  agreement with the theory.
However, the value of the nonlinear attenuation (or saturated SS
intensity) greatly disagrees with the
theoretical  predictions.
The possible reason for the great discrepancy lies in the interplay
between nonuniformity of the SS wave
packet profile due to its generation by FS in the resonance cavity,
destructive interference of the SS waves
coming from different directions on the bolometer, and spatial
inhomogeneity of the bolometer sensitivity.

Another manifestation of the nonlinear SS wave interactions is observation
and
quantitative measurements of the SS spectral broadening with the exponential tails
above
the instability  threshold in the first cell.
The functional dependence of the spectral width on the control parameter and
its
temperature dependence agree well with the  predictions, based on the
kinetic
theory of the 4W  (second order 3W) resonance interactions.
The SS wave system provides a unique opportunity to get the first
experimental evidence of the Gaussian distribution of the amplitudes of the
parametrically generated waves.
This fact supplies a firm ground for an application of the kinetic  theory
with a
random phase approximation to describe statistical behavior of the
parametrically
excited SS wave ensemble.

Interactions of the system of the FS and parametrically generated SS
waves with the SS wave pumped
externally, reveal new effects.
Below the instability threshold, a SS wave, phase conjugated to an
incident SS wave, is excited as a result
of the 3W interaction between the FS pumping field and the SS waves.
Three main features of the phase conjugation, predicted
theoretically, were experimentally verified:
(i) the linear dependence of the amplitude of the conjugated wave on the amplitude
of
the pumping field and on the amplitude of the incident wave,
(ii) the  resonance dependence of the amplitude of the conjugated wave on the
frequency of
the  incident wave, and
(iii) the phase relation between the incident and conjugated waves.
Above the instability threshold two strong nonlinear effects were
observed: the giant amplification of the SS wave
intensity closer to the onset, and the strong resonance oscillations of
the SS wave intensity as a function of
the FS amplitude further from the threshold.
There is no currently quantitative description of these effects.
The qualitative explanation based on the interplay between the (i)
nonuniformity of the SS wave profile, (ii)
incoherence and destructive interference of the SS waves on the
bolometer, (iii) discrete spectra of SS azimuthal
modes in a cell plane, and (iv) nonlinear SS wave interactions and
coherent action of the SS pumping wave,
that cause redistribution of the SS wave intensity between different modes.

Let's now come back to the problem of a possible observation of the fully developed wave
turbulence in
a parametrically excited wave ensemble.
In our experiments we observed the first instability of the  FS wave.
A state of  a large number of the strongly interacting SS waves occurs as a
result
of the parametric instability at $L/\lambda \gg 1$.
The SS wave ensemble has the Gaussian distribution of the wave  amplitudes
that
justifies its description by the kinetic equation with a random phase
approximation.
the 4W interactions generate new modes only in the vicinity of the
main  peak that results in the spectral broadening.
However, this process is not strong enough even far from the  threshold to
redistribute modes in the SS wave spectra in a sufficiently wide wave number
range.
We have not got any evidence of the processes that may lead to creation the
fully developed wave turbulence with a wide frequency spectrum.
However, we cannot exclude a possibility that the fully developed wave turbulence
may be observed at much higher energy fluxes, that is
indeed rather problematic to realize experimentally.
Thus, the parametrically generated SS waves is not suitable system to
study the fully developed wave turbulence contrary to the theoretical expectations
\cite{Pokrov-91,Kolmak-95,Kolmak-95b}.

\section{\bf Acknowledgments.}
\label{sect_Ackn}

We would like to thank V. Cherepanov who was a coauthor of our early publications
(Refs. \cite{Rinb-96,Rinb-97}) and taught us the kinetic theory of the
parametric instability of waves.
We are grateful to A. Muratov whose theoretical guidance on later  stages
helped us in the interpretation of our results.
We are also thankful to V. L'vov, G. Falkovich, and V. Lebedev for many
enlightening discussions and helpful criticism.

This work was partially supported by the Minerva Center for Nonlinear
Physics and
Complex Systems and by the Israel Science Foundation Grant No. 92/96.

\newpage
\begin{figure}
\caption{The vector diagrams for three wave resonance interactions:
               a. Cherenkov process; b. parametric decay process.}
\label{fig_Vect}
\end{figure}

\begin{figure}
\caption{The vector diagram of wave interactions in a resonance cavity.}
\label{fig_ResVect}
\end{figure}

\begin{figure}
\caption{The first cell schematic drawing.}
\label{fig_Cell1}
\end{figure}

\begin{figure}
\caption{Drawing of the flat bolometer.}
\label{fig_BolomFlat}
\end{figure}

\begin{figure}
\caption{Schematic drawing of the second cell. The inset: the fiber
bolometer wiring arrangement.}
\label{fig_Cell2}
\end{figure}

\begin{figure}
\caption{The electronic scheme of the experiment:
          {\bf B} - bolometers (1,2,..8), {\bf H} - heater, {\bf T} -
FS transducer,
          {\bf M} - $ac-dc$ mixer, {\bf G1, G2} - generators, {\bf V}
- voltmeter.}
\label{fig_InstrSet}
\end{figure}

\begin{figure}
\caption{SS wave intensity {\sl vs} FS wave amplitude at fixed temperatures:
          {\bf I}) the first cell for the reduced temperatures
          $\tau = 2.3\times 10^{-4}, 6.52\times 10^{-4}$, and
$1.20\times 10^{-3}$;
          {\bf II}) the second cell for $\tau=3.67\times
10^{-4},6.36\times 10^{-4}$,
           and $8.06\times 10^{-4}$.}
\label{fig_Intens}
\end{figure}

\begin{figure}
\caption{The temperature dependence of the FS threshold amplitude for
the parametric
          instability for: {\bf I}) the first cell, and {\bf II}) the
second cell.
          Solid circles with error bars are experimental points;
          curves {\bf A}-theory for an infinite cell without the spectral spiltting
($r=1,\zeta=1/2$),
          curves {\bf B}-theory for a finite cell with no reflection
from the boundaries and with
          the SS spectral splitting ( $r=0,\zeta=1/\sqrt2$);
          curves {\bf C} (plot {\bf I})-theory for a finite cell with
the reflection coefficient $r=0.7$
          and with the SS spectral splitting ($\zeta=1/\sqrt2$);
          curve {\bf D} (plot {\bf II})-theory of finite cell without
reflections ($r=0$) and
          without the SS spectral splitting ($\zeta=1/2$).
          Dashed lines separate the regions where different conditions
$l/L < > 1$ and
          $\Omega\eta^2/\gamma_2 < > 1$ are satisfied.}
\label{fig_Thresh}
\end{figure}

\begin{figure}
\caption{The SS frequency shift as a function of the reduced
temperature, $\tau$.
          The insets show two types of the SS power spectrum in two
temperature ranges:
          (a) $\tau=2.19\times 10^{-4}$, (b) $\tau=6.52\times 10^{-4}$.}
\label{fig_Spectr1}
\end{figure}

\begin{figure}
\caption{SS spectrum peak shift relatively to $F/2$ {\sl vs} the
reduced temperature, $\tau$,
          just above the threshold of the parametric instability for
different bolometers in the
          second cell.}
\label{fig_FrShift}
\end{figure}

\begin{figure}
\caption{The SS intensity {\sl vs} FS amplitude in the second cell,
measured on the bolometer 4
          at $\tau=3.3\times 10^{-4}$. The dashed line is the linear
fit of the intensity data above
          the threshold $A_{th}$.}
\label{fig_Intens2}
\end{figure}

\begin{figure}
\caption{Temperature dependence of the experimental slope
coefficient, $g_{exp}$, for two
           bolometers 4 and 8, in the second cell.}
\label{fig_Slope}
\end{figure}

\begin{figure}
\caption{SS amplitude fluctuations as a function of time  for
different values of
          $\epsilon=A/A_{th}-1$, at $\tau=2.33\times 10^{-4}$.}
\label{fig_SSampl}
\end{figure}

\begin{figure}
\caption{3D plot of SS power spectral density as a function of FS amplitude
          and SS frequency shift $\delta f=f-\Omega/2$, at
$\tau=2.33\times 10^{-4}$.}
\label{fig_Spect3d}
\end{figure}

\begin{figure}
\caption{SS power spectra for different values of $\epsilon =A/A_{th}-1$, at
          $\tau=2.33\times 10^{-4}$.}
\label{fig_Trian}
\end{figure}

\begin{figure}
\caption{Width of the SS power spectra as a function of the FS amplitude
for different
          temperatures.
          The inset: temperature dependence of the slope, $m$.}
\label{fig_SpWidth}
\end{figure}

\begin{figure}
\caption{Ratio $\Phi=M_{4}/M_{2}^{2}$ for the SS amplitude fluctuations as a
          function of FS amplitude at $\tau=2.33\times 10^{-4}$.}
\label{fig_Stat}
\end{figure}

\begin{figure}
\caption{Amplitudes of incident, $a_1$, (open circles) and conjugated,
$a_2$, (full circles)
          SS waves as a function of FS amplitude $A$ at $\tau=7.07
\times 10^{-4}$ and frequency
          shift $\delta f=-12$ Hz.
          The inset: The spectrum of the SS signal measured on the
bolometer B4 at FS
          amplitude $A=75$ Pa.}
\label{fig_PCMampl}
\end{figure}

\begin{figure}
\caption{Dependence of the phase of the conjugated SS waves, $\phi_2$,
on FS amplitude (open circles),
          and dependence of the sum of phases, $(\phi_1+\phi_2)$, of
the incident and conjugate SS
           waves on the FS amplitude(solid circles).}
\label{fig_PCMphase}
\end{figure}

\begin{figure}
\caption{Frequency dependence of the nonlinear reflection coefficient
at constant FS amplitude
           $A=75$ Pa and  $\tau=7.07 \times 10^{-4}$.
           The dashed line is a fit of the experimental data by a
Lorentzian curve with
           a bandwidth of $\Delta=14$ Hz.}
\label{fig_PCMfreq}
\end{figure}

\begin{figure}
\caption{SS wave intensity {\sl vs} FS amplitude measured by the
bolometer located  opposite
          to the heater that emitted SS wave, in the first cell, for
different heating power
          $w_h$ at $\tau=3.32\times 10^{-4}$.}
\label{fig_SubPump}
\end{figure}

\begin{figure}
\caption{SS wave intensity slope, $g_{exp}$, {\sl vs} SS emitted
intensity, $I_0$, for different
          temperatures, $\tau$.
          The values of $\tau$ are written near the curves.
          The dashed lines are linear fits in logarithmic coordinates:
$g_{exp}\sim I_0^{\nu(\tau)}$.
          The inset: the value of the power index $\nu$ as a function of
$\tau$.
          The $\nu(\tau)$ dependence is linearized in
$\tau$-reciprocal coordinates and fitted by
          the expression: $\nu=\tau_0/\tau-\nu_0$, where
$\tau_0=5.2\times 10^{-4}$ and $\nu_0=1$.}
\label{fig_SubpSl}
\end{figure}

\begin{figure}
\caption{SS wave intensity {\sl vs} FS amplitude measured by the
bolometer located
          opposite to the heater emitted the SS wave in the first cell
for the heater power
          $w_h=108.5$ $ \mu W/cm^2$, and for $\tau=5.21\times 10^{-4}$.}
\label{fig_SubpInt}
\end{figure}

\end{document}